# Structural, Electrical Properties and Dielectric Relaxation in Na$^+$ Ion Conducting Solid Polymer Electrolyte

Anil Arya, A. L. Sharma*

*Centre for Physical Sciences, Central University of Punjab, Bathinda-151001, Punjab, INDIA*

*E-mail: alsharmaiitkgp@gmail.com

**Abstract**

Present paper reports, the structural, microstructural, electrical, dielectric properties and ion dynamics of a sodium based solid polymer electrolyte films comprising of PEO$_8$-NaPF$_6$+ $x$ wt. % SN. The structural and surface morphology properties have been investigated by XRD and FESEM. Different microscopic interaction among the polymer, salt and SN have been analyzed through FTIR. The dielectric permittivity and loss tangent enable us to estimate the dc conductivity, dielectric strength, double layer capacitance, and relaxation time. The loss tangent relaxation peak shifts towards high-frequency side which indicates the decrease of relaxation time and faster ion dynamics. Further, the relation among various relaxation times ($\tau_{\varepsilon'} > \tau_{\tan\delta} > \tau_z > \tau_m$) has been developed systematically. The ionic conductivity has been estimated in whole frequency window including JPL and displays strong dependency on SN content. The sigma representation ($\sigma''$ vs. $\sigma'$) has been introduced for exploring the ion dynamics by highlighting the dispersion region in Cole-Cole plot ($\varepsilon''$ vs. $\varepsilon'$) at lower frequency window and increase in the radius of semicircle indicates a decrease of relaxation time. A well convincing/logical scheme has been proposed which justify the recorded experimental data.

*Keywords:* FTIR, ionic conductivity, dielectric properties, sigma representation, relaxation time

## 1. Introduction

Development of suitable free standing solid polymer electrolyte is the biggest challenge and emerging opportunity for the next generation of high energy density lithium-ion batteries (LIBs) which need to be replaced globally as an alternative source besides the solar and wind energy due to large consumption at both the household and neighborhood level. Polymer electrolytes research attracted the scientific community due to their wide application range such as energy storage/conversion devices, electrochromic devices, supercapacitor, sensors and fuel cells. Amongst them, batteries are playing a more significant role in fulfilling the demand of human beings and industry globally. As the electrolyte is a key component of the battery and is sandwiched between the electrodes. The presence of polymer in the electrolyte system enables it to make a desirable shape, adequate size, with suitable flexibility, mechanical strength, thermal stability, wide electrochemical window, and cost-effective [1-7].

Since the first report (50 years ago) regarding sodium $\beta''$-alumina ($NaAl_{11}O_{17}$) solid state ionic conductor (SSIC) motivated the researchers toward sodium based devices as an alternative to lithium based devices. Sodium ion batteries (SIB) are achieving new sights in the field of energy storage as a promising candidate due to high abundance, low cost and suitable redox potential ($E^o_{Na^+ + Na} = -2.71$ V versus standard hydrogen electrode; only 0.3 V above that of lithium) and lower manufacturing costs as a comparison to the lithium [8-11]. The discovery of Wright & Co-workers in 1973 reported first time ionic conductivity in a polymer matrix comprising of PEO and alkali metals salt. Later, M. Armand recognized their significance in real life such as: energy storage/conversion devices. At present most of the conventional energy storage/conversion devices are based on gel/liquid polymer electrolyte system. The issue of leakage, flammability, hazardous, and poor chemical compatibility prevent the use of liquid electrolytes in high energy storage/conversion devices. Gel polymer electrolytes have high ionic conductivity ($10^{-3}$ S cm$^{-1}$) but the poor mechanical strength due to the presence of plasticized solvents prevents their use as an electrolyte in such applications. At present, solid polymer electrolytes (SPEs) are most auspicious possibility instead of the liquid/gel polymer electrolyte system which use to overcome abovementioned issues. In view of the single ionic conductor of salt with smaller cation and bulky anion plays effective role in fast charge carrier ion dynamics [12].

Poly (ethylene oxide) is most promising aspirant among all polymers since last three decades due to its low glass transition temperature and high degradation temperature. The presence of electron donor ether group ($-\ddot{O}-$) in polymer backbone ($-CH_2 - CH_2 - \ddot{O}-$) makes it interesting for coordination with the available cations. The high value of dielectric constant (~4-5) support the salt dissociation which helps in the release of more free charge carrier's, and hence it may significantly improve the ionic conductivity. But, the semi-crystalline nature of PEO hinders its usage in practical devices. Therefore lowering in crystallinity of host polymer become essential in order to promote the fast transportation of ions via polymer segmental motion [13]. It is well proven that large amorphous phase is a critical requirement as cation migrates only in the amorphous system. Here, in the amorphous polymers, ion dynamics is achieved by the elimination and formation of coordinating sites for the cation. Now, during the jump of an ion from one coordinating site to other, previous site use to be relaxed and the creation of new site occurs which adopts the coming ion. The coupled motion of the ion with PEO chain having ether group promotes the ion migration [14-15]. So many attempts have been adopted for the enhancement of the ionic conductivity and the polymer flexibility of the system. One effective approach is the dispersion of the nanofiller such as $TiO_2$, $SiO_2$, $CeO_2$, $BaTiO_3$ etc. which improves the ionic conductivity by providing the additional coordinating sites to the cation. The Lewis acid-base interaction benefits objectively in the effective role of nanofiller for enhancement of the ionic conductivity [16-21]. Another approach is the use of the high dielectric constant solvents such as: EC, PC, DEC and DMC that reduces the polymer recrystallization tendency and supports the polymer chain sliding which leads to high ionic conductivity owing to the enhanced polymer flexibility [22-25].

Plastic crystals have recently stated as a versatile candidate for the SSIC due to their plastic nature which makes them capable of deformation without fracture under applied stress. Also, the free charges carriers are vital for a fast ionic conductor which directly depends on the dielectric constant of plastic and hence the energy storage capacity (energy density $\propto$ dielectric constant). Such component with the high dielectric constant is the best clue since it favors the target of achieving fast ion transport and controls the overall separation of charge (cation and anion) or polarization. Succinonitrile (SN; N≡C–$CH_2$–$CH_2$–C≡N) is the most prominent candidate in order to improve desirable properties of SSIC due to its high polarity, dielectric constant (~55), and adequate melting point (62 °C), which supports more dissociation of salt in the polymer [26]. The above properties motivated us to study the effect of Succinonitrile (SN) on the dielectric relaxation and ion dynamics of prepared polymer salt complex film. Since dielectric parameters play a valuable role in the polymer electrolytes in terms of energy storage as well as in electrical transport process of involved charge carrier. As stated earlier that the relaxation of one coordinating site promotes the ion migration to another coordinating site via ether group (electron rich group) of the host polymer. Hence, the shorter relaxation time seems to be favorable for rapid ion transport. It can be represented by the relationship, $\sigma\tau T = constant$, in the investigated polymer electrolyte system. So, the frequency dependent dielectric parameters such as complex conductivity, complex permittivity and relaxation time provide valuable information regarding the ion dynamics of the solid state ionic conductors (SSIC) and are important to understand the ion dynamics along with relaxation [27-31]. The various relaxation time has been evaluated through different relevant plot (i.e. dielectric loss, tangent loss, modulus and imaginary part of the impedance) systematically and correlated fairly [26, 32-35].

The present investigation was accompanied to study the role of SN in improving the ion migration in prepared polymer electrolyte system and to understand the ion dynamics over the whole frequency window. X-ray diffraction and Fourier transform infrared spectroscopy are used to probe the complex formation and visualizing polymer-ion, ion-ion interactions respectively. Impedance spectroscopy was used to study the electrical conductivity results and ion transport number analysis. As the number of free charge carriers directly influence the dielectric constant of the system. Therefore, the role of SN incorporation on various dielectric parameters ($\varepsilon'$, $\varepsilon''$, $\sigma_{ac}$, $\tan \delta$) has been explored. All dielectric parameters were simulated using the concerned equation that validates the present experimental study. The various transport parameters such as number density of charge carriers, mobility, diffusion coefficient and relaxation time (average & molecular) are properly correlated with the ionic conductivity results. We have proposed a novel scheme to highlight the interaction between the constituents of the solid polymer electrolyte matrix. The molecular structures of PEO, NaPF$_6$ and SN are shown in figure 1.

## 2. Methodology

### 2.1. Materials Preparation

PEO with a molecular weight of $1\times10^6$ g/mol (Sigma-Aldrich), NaPF$_6$ and SN from Aldrich were used as base material. Anhydrous acetonitrile from Sigma Aldrich was used as a solvent. The host polymer (PEO) and the salt NaPF$_6$ were vacuum dried before use. The Ö/Na$^+$ ratio was constant 8:1 that is an optimized value for high ionic conductivity. The polymer electrolytes were synthesized via standard solution cast technique. Initially, the 0.5 g of PEO and appropriate salt (Ö/Na$^+$=8:1) were dissolved in the 15 mL acetonitrile (ACN) at room temperature for 6-8 hr. Then varied wt. % of SN was added to the polymer salt solution and again stirring was done for 18-20 hr. The resulting viscous solution was casted on the Petri dishes and stored in a desiccator with silica gel till a dry film is obtained. Further drying is done in the vacuum oven to remove the residual solvent. The samples are designated as SPE-$x$ where $x$ =1, 2, 3, 4, 5, 6 & 7. SPE 1 designates the pure PEO and SPE 2 to SPE 7 are designated for the wt. % of SN added (0, 1, 2, 3, 4, 5 wt. %) PEO$_8$-NaPF$_6$ sample.

### 2.2. Characterization

X-ray diffraction (XRD) (Bruker D8 Advance) was performed for the determination of crystallinity and recorded with Cu-K$_α$ radiation (λ=1.54 Å) in the Braggs angle range (2θ) from 10° to 60°. The Fourier transform infra-red (FTIR) spectra (Bruker Tensor 27, Model: NEXUS–870) were recorded in absorbance mode over the wavenumber region from 600 to 3500 cm$^{-1}$ to probe the presence of various interactions such as polymer-ion, ion-ion interaction and complex formation. The ionic conductivity was measured by complex impedance spectroscopy (CIS) in the frequency range of 1 Hz to 1 MHz using the CHI 760 electrochemical analyzer. An AC sinusoidal signal of 10 mV is applied to the cell configuration *SS/SPE/SS* where solid polymer electrolytes films are sandwiched between two stainless steel electrodes. The ion transference number was measured by i-t characteristics by applying a voltage of 10 mV. As, the present report is focused on the dielectric relaxation and charge carrier dynamics, so it becomes important to go for the dielectric analysis. Then the impedance data is transformed into the dielectric constant, dielectric loss, complex conductivity data which is further transformed into the real and imaginary part of the modulus. All plots were fitted with corresponding equations by Origin® 8 software to evaluate the various parameters that enable us to explore the ion dynamics.

## 3. Results and Discussion
### 3.1. X-ray diffraction (XRD) analysis

To investigate the structural investigations of the prepared polymer electrolyte, XRD is a crucial technique which provides details in terms of the shift in the peak position, peak broadening, and lowering in peak intensity. Figure 2 depicts the XRD diffractograms of the pure PEO (figure 2 a) and polymer salt complex (fig 2 b) with varying SN content (figure 2 c-g) over the 2θ range 10º to 60º. From the figure it is noted that the pure PEO shows the two characteristics diffraction peaks at 19º and 23º which confirms the semi-crystalline nature of the PEO, corresponding to the planes 120 and 112/032 [36]. Some additional minor peak (24º and 25º) are observed close to characteristics peak and may be due to the cation coordination with electron rich group of PEO [37]. The NaPF$_6$ exhibits sharp and intense diffraction peaks near 19º and 23º along with two minor peaks near 32º. The absence of salt peak indicates the complete dissociation of the salt. The d-spacing was obtained using the Bragg's formula 2dsinθ=nλ and interchain separation (R) using the equation R=5λ/8sinθ and determined values are shown in Table 1. Further addition of SN shifts the peak toward lower angle and decrease in intensity is observed [37, 38]. This evidences the noticeable increase of the interplanar spacing (d-spacing) and interchain separation (R) which may be due to penetration of anion in between the polymer chain (Table 1). Pure SN exhibits the sharp characteristic peaks at 20° and 28° and no such diffraction signature is observed in the present system, which indicates that polymer, salt, and SN are well dissolved in the polymer system evidencing the complex formation [39]. The decrease in intensity is more for the SPE 6 system with 4 wt. % SN (fig. 2 f) and indicates the suppression of the crystalline phase due to penetration of SN in between the polymer chains. This increases the interchain separation which indicates that more free volume is available for conduction (Table 1). The amorphous content is increased on addition of SN due to the disordered in the polymer chain and hence the faster ionic transport is supposed to occur [14].

### 3.2. Field Emission Scanning Electron Microscopy (FESEM) Analysis

The FESEM micrographs of the host polymer and polymer salt complex are depicted in the Figure 3 a-b, respectively. While the micrograph for the SN incorporated polymer complex are shown in the Figure 3 c-f. From the micrograph, it is observed that the pure PEO shows its semi-crystalline nature and rough surface evidence the same. The presence of pore may be due to fast evaporation of the solvent in solution cast technique. After addition of the salt, surface morphology gets modified and may be attributed to the interaction of ether group of polymer with the cation as evidenced by XRD. Now, the addition of SN in the polymer salt complex smoothens the surface and indicates the elimination of the crystalline phase. The formation of the smoother surface after addition of the SN evidences the increase of amorphous content owing to the interaction between the constituents of the present system. It is observed from the micrographs that the smoother morphology is superior in the Figure 3 e. It indicates that the smoother morphology must be accompanied by the maximum ionic conductivity and transport parameters which are to be analyzed in the upcoming sections. At high SN content, there is not sufficient dissolution of the SN in the polymer salt system and plastic nature of SN dominates which can be seen in the micrograph (Figure 3 f). It can be concluded that the incorporation of SN in the polymer salt system enhances the amorphous content which is desirable for the fast solid state ionic conductor.

### 3.3. Fourier Transform Infra-red Spectroscopy (FTIR) Analysis

The FTIR is a powerful tool to probe the possible polymer-ion, ion-ion interaction and spectral changes observed in the functional group of polymer host due to the addition of salt and SN. The FTIR absorbance spectra of the host polymer and polymer salt complex with $x$ wt. % (x=0, 1, 2, 3, 4 and 5) SN is depicted in figure 4. The various absorption bands corresponding to modes of PEO are observed and complexation of polymer with salt is confirmed. The peak located at 950 cm$^{-1}$ is corresponding to the C–O stretching vibration mode and shifts towards lower wavenumber side suggesting the effect of salt on host polymer. The fundamental peak for PEO is observed near 1100 cm$^{-1}$ and is interesting for us since it expresses the interaction of cation with ether group of the polymer chain and corresponds to symmetric and asymmetric C-O-C stretching mode. The change in peak intensity and position indicates the interaction of cation (Na$^+$) with an ether group ($-\ddot{O}-$) of host polymer [37, 40]. Further addition of SN results in asymmetry of peak suggesting that the SN plays an effective role in altering the interaction present in the polymer salt complex. The CH$_2$ bending and asymmetric/symmetric twisting vibrations are observed near 1340 cm$^{-1}$, 1282 cm$^{-1}$, and 1230 cm$^{-1}$ respectively. All solid polymer electrolyte films show strong absorption band in the region 2800 cm$^{-1}$ to 2950 cm$^{-1}$ which correspond to the symmetric and asymmetric vibration of C-H stretching mode of CH$_2$ group in PEO [36]. Further addition of SN alters the stretching mode (peak position and shape) and two separate stretching modes are clearly visible now. The strong evidence of interaction is obtained which confirm the complex formation of the polymer, salt, and addition of SN further modify the polymer chain arrangement.

Since the Na$^+$ is IR inactive therefore our focus become essential on the evaluation of anion peak instead of cation and reflection appearing in anion peak are transformed into cation for the analysis. In the free anion peak area, the peak position is attributed to the interaction between polymer and PF$_6^-$ anion and is evidenced by deconvoluting the hexafluorophosphate (PF$_6^-$) anion band confirming the presence of free ions and ion pairs. The typical vibrational mode of PF$_6^-$ anion in SPEs films have been observed in wavenumber region ~800-900 cm$^{-1}$. The presence of asymmetry in the peak for all SPE system may be attributed to loss in degeneracy from octahedral symmetry O$_h$ →C$_{3v}$ arising due to the simultaneous presence of more than two components i.e. free ion, ion pairs. So to study above degeneracy and presence of both ions, deconvolution of PF$_6^-$ peak is done as explained somewhere [41-42].

As the number of free ions is an important parameter in the polymer electrolytes as it affects directly the ionic conductivity. So, here we evaluate the free ions and ion pair contribution using equation 1:

$$\begin{cases} \text{Fraction of free anion} = \dfrac{\text{Area of free ion peak}}{\text{Total peak area}} \\ \qquad\qquad\qquad and \\ \text{Fraction of ion pair} = \dfrac{\text{Area of ion pair peak}}{\text{Total peak area}} \end{cases} \qquad (1)$$

The deconvoluted pattern of SPE films is shown in Figure 5 and the peak at lower wavenumber is attributed to free ions and at higher wavenumber due to ion pairs. The changing profile (change in position and area) of free ion and ion pair peak on the addition of the SN illustrates the change in polymer-ion interaction due to the presence of polar nitrile group in SN [43]. A relative comparison of corresponding free ion area and ion pair area is summarized in Table 2 and high free ion area obtained for the SPE 6 with 4 wt. % SN. The increase in area may be due to the release of more number of free charge carriers because of high polarity of nitrile group present in the SN. The increase in free ion area displays a similar trend as shown by the ionic conductivity and is correlated in the next section.

### 3.4. Impedance Spectroscopy Analysis

The impedance analysis of the SPE films was carried out at room temperature by sandwiching them between two stainless steel (SS) ion blocking electrodes with a voltage signal of ~ 20 mV. The log-log plot between imaginary vs. real part ($Z''$ vs. $Z'$) of impedance is displayed in Figure 6. It was chosen over the traditional plot for better clarity and comparison of the different impedance plots in a single plot of unique way. The enthusiasm behind this representation was from the superiority of log plots as elaborated by Jonscher [44-45]. In obtained experimental results all the polymer electrolyte system shows identical feature comprising of small semicircular nature at high frequencies followed by an arc at lower frequency region. The semicircle nature is attributed to the ionic conducting nature of prepared polymer electrolytes while the suppressed arc at lower frequency corresponds to the double layer capacitance effect due to blocking of ions at the end of blocking stainless steel electrodes. The dip in the graph is corresponding to minima in $Z''$ on x-axis toward high frequency corresponds to the bulk resistance of sample and lowering of bulk resistance corresponds to the high ionic conductivity of polymer electrolytes [32, 46].

It is observed that for pure polymer electrolyte (Figure 6) the dip in the arc is at large value on the x-axis hence associated with the bulk resistance in the polymer electrolyte system. Now, when the salt is added to the polymer host then the dip in the curve shifts toward the lower impedance value, which indicates lower bulk resistance and hence the enhanced ionic conductivity. Then the addition of even a small amount of SN in the polymer salt matrix depicts the same pattern with decreased value of bulk resistance and it indicates the triggered ion transport in the investigated polymer electrolyte system. The ionic conductivity of the SPE film depends on the concentration of the mobile ions, ion charge, and mobility (how easily an ion is moved through the solid) and expressed as; $\sigma = ne\mu$. The Ionic conductivity of all SPEs was calculated by using the equation 2:

$$\sigma = \frac{t}{R_b \times A} \quad (2)$$

Where '$t$' is the thickness of the polymer electrolyte (cm), 'A' is the area of the blocking electrode (cm$^2$) and $R_b$ is the bulk resistance of polymer electrolyte films.

For, the sample (4 wt. % SN) the highest ionic conductivity is due to the lowest value of bulk resistance as shown in the inset. The host polymer shows a lower value of ionic conductivity and addition of salt in the host polymer system shows enhancement in conductivity. The increase of ionic conductivity is attributed to the formation of free charge carriers via the interaction of cation with electron-rich ether group of the host polymer. The coordinating sites provided by the polymer chain supports the migration of cation which lead to the faster ion conducting system. Further, the addition of SN dissociates more salt due to high polarity of SN (N≡C–CH$_2$–CH$_2$–C≡N) molecules and makes polymer chain flexible which leads to the faster segmental motion of polymer chains. The highest ionic conductivity is obtained for the SPE 6 which contains 4 wt. % SN in the polymer salt system. The low value of bulk resistance ($R_b$) for SPE 6 as compared to other samples evidences the high ionic conductivity of the same. The decrease of ionic conductivity at high SN content may be due to the dominance of plastic effect of SN which leads to ion pair formation or leads to blockage of ion conducting pathways. The double layer capacitance observed at low-frequency side is calculated using the equation 3 and given in table 3:

$$C_{dl} = -\frac{1}{\omega Z''} \quad (3)$$

Where $\omega$ is the angular frequency and $Z''$ is the imaginary part of impedance at low frequency. The high value of double layer capacitance is achieved for the highest ionic conducting sample and it is in agreement with the deconvoluted FTIR results which concluded the highest free ion area for the same. The correlation of free ion area, ionic conductivity and double layer capacitance has been built by plotting the graph in the last section of the manuscript. Since, the ions are playing an important role in the current system so ionic nature of polymer electrolyte needs to be evidenced as reported in the subsequent section.

### 3.5. Ion transference number

The ion transference number in polymer electrolyte system is basically the fraction of the total current carried by the respective ion across a given medium to the total current. Figure 7 shows the variation of polarization current as a function of time for the pure polymer and polymer electrolyte with SN content at room temperature [47]. Almost, an identical pattern is observed for all SPEs, which shows a very high initial total current ($I_t$) followed by a sharp drop in its value with the passage of time. After some time current get saturated with time, the initial high current is attributed to the contribution from both the ions and the electrons ($I_t = I_e + I_i$) while the final saturation/residual current after polarization is attributed to the contribution of electrons only due to ion blocking electrodes (SS) at the interface. For all SPE system ionic transference number was about ~0.99 and it reveals the ionic conducting nature of polymer electrolyte [48]. The ionic and electronic conductivity were evaluated and recorded in table 3 using the equation 4:

$$\begin{cases} i_t = i_i + i_e \\ \sigma_{ionic} = \sigma_{electrical} \times i_i \\ \sigma_{electronic} = \sigma_{electrical} \times i_e \end{cases} \quad (4)$$

The ionic conductivity and electronic conductivity are also in correlation with electrical conductivity value. The negligible value of electronic conductivity justifies the use of prepared electrolyte as a separator cum electrolyte in the present existing energy conversion devices.

### 3.6. Dielectric Spectroscopy Analysis

### 3.6.1. Cole-Cole plot

The Cole-Cole plot ($\varepsilon''$ $vs.$ $\varepsilon'$) is a powerful tool for materials retaining one or more well separated relaxation processes with comparable magnitudes and obeying the Cole-Cole formalism. The plot comprises of variation of dielectric loss with dielectric storage component at constant temperature and formation of perfect semicircle indicates the presence of single relaxation time. Figure 8 shows a depressed semicircular arc due to broad relaxation region and maximum loss occurs at the midpoint of the semicircle (solid red lines depict the fitted plot), while the dielectric values are read from right to left with an increase of frequency. One interesting conclusion is observed, that at a dielectric constant of infinite frequency ($\epsilon_\infty$) and static dielectric constant ($\epsilon_s$) there will be no loss [49]. All SPEs shows a semicircle arc but the SPE system with 4 wt. % SN displays a tail at low frequency which may be due to high loss. Here, charge carriers act slowly under the applied electric field and dispersion is observed while at high frequency dielectric loss decreases continuously and approaches to zero.

### 3.6.2. Dielectric spectrum analysis

The dielectric storage and dielectric loss are two important properties for determining the suitability of polymer electrolyte for energy storage applications. Dielectric analysis of solid polymer electrolyte materials is desirable to attain the better insight into ion dynamics and are analyzed briefly by highlighting in terms of the real and imaginary parts of complex permittivity ($\varepsilon^*$) [50-52]. The dielectric permittivity describes the polarizing ability of a material in the presence of an external electric field and is a function of frequency represented by equation 5:

$$\varepsilon^* = \varepsilon' - j\varepsilon''\ ;\ \varepsilon' = \frac{-Z''}{\omega C_o(Z'^2 + Z''^2)}\ \text{and}\ \varepsilon'' = \frac{Z'}{\omega C_o(Z'^2 + Z''^2)} \quad (5)$$

Where $\varepsilon'$ and $\varepsilon''$ represent the real and imaginary parts of the dielectric permittivity and j the imaginary unity ($j^2 = -1$). The real part of dielectric permittivity ($\varepsilon'$) is proportional to the capacitance and measures the alignment of dipoles or polarization, whereas the imaginary part of dielectric permittivity ($\varepsilon''$) is dielectric loss and is proportional to conductance and represents the energy required to align the dipoles. Cole-Cole proposed the distribution of relaxation time for Debye processes [53], given by equation 6:

$$\varepsilon_* = \epsilon_\infty + \frac{\Delta\varepsilon}{1 + (jx)^{1-\alpha}} \quad 0 \leq \alpha < 1 \quad (6)$$

Where, $\alpha$ is distribution parameter and $x = \omega\tau$ ; $\omega$ is the angular frequency of applied field and $\tau$ is Debye relaxation time. The real and imaginary parts of dielectric constant can be obtained by separating above equation and expressed as equation 7 a & b [54];

$$\varepsilon' = \epsilon_\infty + \frac{\Delta\varepsilon(1 + x^{1-\alpha}\sin \pi/2\alpha)}{1 + 2x^{1-\alpha}\sin \pi/2\alpha + x^{2(1-\alpha)}} \quad (7a)$$

$$\varepsilon'' = \Delta\varepsilon \frac{x^{1-\alpha}\cos \pi/2\alpha)}{1 + 2x^{1-\alpha}\sin \pi/2\alpha + x^{2(1-\alpha)}} \quad (7b)$$

Above equations can be written in another form by replacing $\alpha$ with $1 - \alpha$ in equation 7 and we obtain equation 8 a & b [55].

$$\varepsilon' = \epsilon_\infty + \frac{\Delta\varepsilon\left(1 + x^\alpha \cos \frac{\alpha\pi}{2}\right)}{1 + 2x^\alpha \cos \frac{\alpha\pi}{2} + x^{2\alpha}} \quad (8\ a)$$

$$\varepsilon'' = \Delta\varepsilon \frac{x^\alpha \sin\frac{\alpha\pi}{2}}{1 + 2x^\alpha \cos\frac{\alpha\pi}{2} + x^{2\alpha}} \quad (8\,b)$$

Here, $\varepsilon_s$ is static dielectric constant ($x \to 0$), $\varepsilon_\infty$ is dielectric constant ($x \to \infty$), $x = \omega\tau$; $\omega$ is the angular frequency of applied field and $\tau$ is Debye relaxation time (reciprocal of jump frequency in the absence of external electric field). Here, $\alpha$ is distribution (power law) exponent of the material sample. The fitted parameters are shown in Table 4 at RT. From the Table 4 it is observed that decrease in value of $\alpha$ is a direct indication of more distributed relaxation time. The value of $\varepsilon_\infty$ is maximum for the SPE 6 system and corresponds to 4 wt. % of SN content as shown in Table 4.

Figure 9 shows the plot of frequency dependence of dielectric constant ($\varepsilon'$) for different SN content at RT and an absolute agreement is obtained between the experimental & fitted data over the measured frequency range. All SPEs shows the strong frequency dispersion at a lower frequency followed by a frequency independent region above 1 kHz. The addition of SN in the polymer salt system provides suitable evidence of the effective role played by SN in polymer salt matrix. The high value of dielectric constant at a lower frequency in the graph indicates dielectric polarization and the same trend is observed in all polymer electrolytes. Generally, there are two sources of dipoles present in polymer electrolytes. One is due to dissociation of salt (cations and anions) and another is because of ether group of PEO. Former one provides a free number of charge carriers for transport. When an electric field is applied these charges migrate along the field appropriately, but the blocking electrodes present in the circuit prevent transport of ions via an external circuit. The accumulated ions forms a polarization region due to an increase of charge species. The latter one deals with the significant contribution in the electron rich ether group ($-\ddot{O}-$) of host polymer chain (PEO). When an electric field is applied then there may be conformational changes in the polymer chain and it may lead to a nonzero contribution to polarization. These both contributions simultaneously lead to the high value of the dielectric constant.

At low frequency, the high value of dielectric constant may be due to the presence of ion pairs which are enabled to do long-range migration and behave like localized dipoles in the immobilized state. These localized dipoles respond to the externally applied electric field due to sufficient time which increases the dielectric constant and the bulk capacitance [56]. Also, it is evidenced that dielectric constant increases with the addition of SN and may be due to better dissociation of salt due to the presence of polar nitrile group in SN. The highest value of dielectric constant at all frequency provides direct evidence for highest conducting sample and is also in correlation with FTIR data. The decrease of dielectric constant at high SN content may be due to the formation of ion pairs via increased coulombic interaction or insufficient active SN content. The decrease of free charge carriers decreases the accumulation of charges and hence, the dielectric constant. Now, in the high-frequency window decrease in dielectric constant is observed for all samples due to relaxation processes. This decrease may be due to the inability of heavy positive and negative charge carriers to rotate/translate along the filed direction and this directly affects the dielectric constant value. Now, the field changes direction before the dipoles get aligned along the field. The dielectric constant is now frequency independent as the fast periodic reversal of the electric field prevents the ion diffusion in the direction of the field. The constant region is achieved due to the failure of molecule dipoles to follow the field and contribution to dielectric constant due to orientation process ceases. Also the increase of dielectric relaxation strength ($\Delta\varepsilon = \varepsilon_s - \varepsilon_\infty$) with SN is a measure of increased ionic polarization. From Table 4, it is clear that value of $\Delta\varepsilon$ is higher for the SPE 6 system and relaxation time is lower which revels the batter dissociation of salt as compared to other SN content. We have also calculated the molecular relaxation time and that is also in one-to-one agreement with the impedance and FTIR study [57]. The lowest value of molecular relaxation time is for the 4 wt. % SN and evidences the enhancement in ionic conductivity as obtained from electrical conductivity analysis [See Supplementary discussion and Figure S 1]. Further, this supports the impedance study which shows the highest ionic conductivity for the same concentration and directly evidences the increased number of free charge carriers [36].

As dielectric constant varies with the frequency of applied field so a comparison of dielectric constant at different frequencies is depicted in Figure 10 and the dielectric constant is higher at lower frequencies and decreases with the increase of frequency. One fascinating point to be noted is that the dielectric constant is higher for the highest conducting sample as expected from FTIR and Impedance results. As the high value of dielectric constant indicates the better dissociation of salt which directly indicates the more free charge for migration owing to the formation of the conductive network inside polymer matrix [58].

The fitted plot of dielectric loss ($\varepsilon''$) versus frequency (using Eq. 8 d) is shown in Figure 11 and follows the same trend as real part of permittivity. A close agreement is observed between the experimental and fitted curve and dielectric loss decreases with increase of the frequency. The ions present in the system shows the effect of inertia on the application of the field. But in the high frequency window, when the periodic reversal field occurs then a three-step process occurs. First, the de-acceleration of ions occurs and then ion stop for a negligible time and at final stage ions get accelerated in the reverse direction. This generates some internal heat in dielectric and called as dielectric energy loss ($\varepsilon'' = 0$ for $\omega\tau = 0$). The dielectric loss at low frequency is mainly due to dc resistivity while, at high frequencies is due to dipole rotations from low to high energy states. As the addition of SN helps in dissociation of salt and in release of number of charge carriers which leads to large heat generation or large dielectric loss [59]. The absence of relaxation peak in all graphs may be due to large electrode polarization effect which masks the relaxation behavior of polymer electrolytes. Therefore, electric modulus and ac conductivity are attempted as alternatives to study ionic dynamics in the next sections.

### 3.6.3. Tangent delta analysis

The loss tangent (tan δ) is defined as the ratio of energy loss to energy stored in a periodical field and is also called as dissipation factor. The component in phase with applied voltage results in loss and δ is the loss angle. A maximum in tangent delta vs. frequency is obtained for a particular combination of frequencies which satisfy the equation $\omega\tau = 1$; where ω is the angular frequency of applied field and $\tau$ is Debye relaxation time (reciprocal of jump frequency in the absence of external electric field). The loss is maximum for this particular frequency as explained below. As a molecule of a dielectric possesses all translational, vibrational, and rotational/orientation motion in absence of field (E=0). When an alternating electric field is applied (E≠0) then a constraint is sensed by the molecules or dipoles in changing their directions along the field only when frequency of applied external electric field and frequency of molecule rotation matches properly. This matching results in maximum power transfer to the molecular dipoles from applied filed and heating is produced in the system [60].

     Figure 12 shows the variation of loss tangent with frequency for different solid polymer electrolytes and the signature of single relaxation peak is a fingerprint of ionic conduction via polymer chain segmental motion [61]. Initially, at lower frequency region, an increase of loss tangent may be attributed to the dominance of Ohmic (active) component than capacitive component (reactive). But the inverse trend is observed with the increase in frequency and this decrease of loss tangent owing to the independent nature of Ohmic part and growth of reactive

component with frequency. The presence of resonance peak is also in good agreement with the theoretical approach proposed by Debye and corresponds to the maximum transfer of energy on the application of field [62].

Now, the addition of salt in the polymer host shifts the relaxation peak towards the high-frequency side which indicates the decrease of Debye relaxation time. The change in the loss peak and a shift in position is evidence of the presence of dielectric relaxation processes and a decrease of relaxation time [63]. The decrease of relaxation time evidences the increase of ionic conductivity and is in agreement with the FTIR data. To obtain the relaxation time the tan δ vs. frequency plot is fitted generally with equation 9:

$$\tan \delta = \frac{(r-1)}{r+x^2}x \qquad (9)$$

Where r is the relaxation ratio ($\varepsilon_s/\varepsilon_\infty$), $\varepsilon_s$ is static dielectric constant ($x \to 0$), $\varepsilon_\infty$ is dielectric constant ($x \to \infty$), $x = \omega\tau$; $\omega$ is the angular frequency of applied field and $\tau$ is Debye relaxation time (reciprocal of jump frequency in the absence of external electric field). This equation proposed by Debye provides us satisfactory fitting for a single particle and non-interacting system (null interaction between dipoles). But, in the low frequency window the Debye model is not followed properly and may be due to the presence of multi type dipole polarization or heterogeneous type complex nature of the material. In the present investigated material system, broad loss peak is observed and experimentally evidenced broad peak motivates us to do modification in ideal Debye equation for better simulation of experimental results. As, in present case polymer matrix is an interacting system as evidenced by FTIR and this many body interaction need to be studied deeply by fitting of loss plot satisfactorily. Therefore, in order to meet the experimental needs it becomes essential to do certain empirical modification by adding some parameter as one parameter is used in Cole-Cole, Davidson-Cole, Williams-Wats, and two parameters in Havriliak-Negami fluctuations [64]. So, in this ideal Debye equation shape parameter α is added as the power law exponent with value $0 \leq \alpha \leq 1$ to fit the broad tangent delta plot (equation 10). This proposed empirical equation and presence of this factor confirms the presence of more strong interaction in our system. The modified equation has been written as equation 10:

$$\tan \delta = \left(\frac{(r-1)}{r+x^2}x\right)^\alpha \qquad (10)$$

This equation was used for calculating the fitting parameters in tangent delta plot and the fitted parameters are recorded in Table 5. This equation provides the empirical confirmation of our results and also the fitting observed in Figure 12 displays a close agreement of measured and fitted results. The fitted curve is represented by the solid red lines and describes the observed results with accuracy. The negligible slight deviation at low-frequency side probably may be due to the electrode polarization or the diffusion of ions towards the electrodes [65]. We can return to Debye model for $\alpha = 1$ in equation 10. The physical significance of $\alpha$ is not yet been worked out but to be done in future and it will provide us crucial aspects which will justify the proposed equation.

The decrease of relaxation time with the addition of SN is due to increase of chain flexibility by the interaction of cation and polymer chain. This leads to increase in amorphous content and more free volume is available for ion migration and hence a reduction in the relaxation time is observed. The lowest value of relaxation time is observed for the SPE 6 (4 wt. % SN) and the same system shows the highest value of free ion area and ionic conductivity value. So, the overall effect of SN is a reduction of relaxation time and enhanced ionic conductivity. While at very high SN content insulating nature of plasticizer SN appears that causes the increase of relaxation time in the system. The lowest value of relaxation time for SPE6 provides direct evidence for the faster ion dynamics in plasticized polymer electrolyte and supports the FTIR as well as impedance data. Therefore, it may be concluded that modified equation may be useful for describing the highly interacting systems and issue of fitting at a lower frequency can be resolved. In order to investigate the dielectric analysis fully, the suppressed features of Cole-Cole plot at high frequencies are explored with a new approach *first time in case of solid state ionic conductors (SSICs)* in the upcoming section.

### 3.7. Sigma representation ($\sigma''$ vs. $\sigma'$)

Cole-Cole plot ($\varepsilon''\ vs.\ \varepsilon'$) is a very useful tool when the material possesses relaxation processes and obey Debye equations. However, when the material possesses high ionic conductivity, the Cole-Cole representation becomes less useful, because the presence of dc conductivity leads to a divergence of $\varepsilon''$ at lower frequencies. Therefore, a new approach ($\sigma$-representation ($\sigma''\ vs.\ \sigma'$) has been attempted to describe the divergence in Cole-Cole plot [66]. This new representation is the best alternative for getting insights of the high frequency data and is capable of eliminating the issue of high frequency data in Cole-Cole plot. One extremely crucial point is that since both $\sigma''\ and\ \sigma'$ involve the multiplication of frequency with real and imaginary dielectric parameters, as a consequence high frequency features turn out to be noticeable in $\sigma''$ vs. $\sigma'$ plot which seems to be suppressed in the Cole-Cole plot.

The complex electrical conductivity can be written using following expression (equation 11)

$$\begin{cases} \sigma(\omega) = \sigma' + i\sigma'' & (a) \\ \sigma_\infty = \sigma_o + \frac{\varepsilon_v(\varepsilon_o - \varepsilon_\infty)}{\tau} = \sigma_o + \delta & (b) \\ \sigma_{ac} = \sigma' = \omega\varepsilon_v\varepsilon'' \text{ and } \sigma_{dc} = \sigma'' = \omega\varepsilon_v(\varepsilon' - \varepsilon_\infty) = \omega\varepsilon_v\varepsilon' & (c) \\ r = \frac{\delta}{2} = \frac{\varepsilon_v(\varepsilon_o - \varepsilon_\infty)}{2\tau} & (d) \end{cases} \quad (11)$$

Here, $\sigma'$ is a real part of conductivity, $\sigma''$ is imaginary part of conductivity, $\omega$ is the angular frequency, $r$ is the radius of the semicircle. It is very clear that, when $\sigma'' = 0$ then low frequency x-intercept gives dc conductivity ($\sigma_o$) and high frequency x-intercept gives $\sigma_\infty$. The radius of the semi-circle (r) is inversely proportional to the relaxation time ($\tau$). Figure 13 shows the plot of $\sigma''$ vs $\sigma'$ and all SPEs shows a clear semicircle which confirms that material is obeying the Debye conducting feature. The low frequency intercept and high frequency intercept provides us $\sigma_o$ and $\sigma_\infty$ respectively. Table 6 shows the various parameters obtained from the above plot and increase of dc conductivity is evidence with the addition of SN with maxima for SPE 6.

Also, inset of figure 13 shows the increase of radius of a semicircle with the addition of SN in polymer salt system which indicates the decrease of relaxation time and is evident of faster ion dynamics. As 'r' is inversely proportional to $\tau$, so the highest value of r for SPE 6 provide strong evidence for fast ion migration and is also in good agreement with the impedance, FTIR and dielectric study. At frequency much higher than the characteristic frequency, $\sigma''$ vs. $\sigma'$ plot shows the straight line which is due to $\omega^{-\alpha}$ type asymptotic behavior of $\varepsilon''$ and $\varepsilon'$. The simulated plot of Sigma representation is shown in supplement file with explanation (Figure S2).

### 3.7.1. AC conductivity analysis

The AC electrical measurements (AC conductivity) of all plasticized polymer electrolyte has been obtained using the equation 12:

$$\sigma' = \sigma_{ac} = \omega\varepsilon_o\varepsilon'' = \omega\varepsilon_o\varepsilon' \tan\delta \quad (12)$$

Where $\omega$ is the angular frequency, $\varepsilon_o$ is the dielectric permittivity of the free space and $\varepsilon''$ represents the dielectric loss. Figure 13 a shows variation of ac conductivity against the frequency for different wt. % of SN in polymer salt matrix. The frequency dependent real part of electrical conductivity indicates that there are three distinct regions, (i) low frequency dispersion region, (ii) frequency independent plateau region and, (iii) high frequency dispersive region.

The low-conductivity value at the low-frequency dispersion region evidences the accumulation of ions (electrode polarization) due to the slow periodic reversal of the electric field and disappears with an increase in frequency. The intermediate constant conductivity region at the slightly higher frequency is the result of long-range diffusion of ions and dc conductivity can be extracted from it. Region (iii) is due to short range ion transport associated with ac conductivity (hopping of charge carriers). All SPE system shows a shift in both intermediate frequency region and high-frequency region towards high frequency. For the highly conductive system, a small high-frequency dispersion region evidences the ion migration via hopping.

The conductivity of pure polymer is very low due to EP effect and addition of salt increases the ionic conductivity. Two types of phenomena occurs (i) long-range conduction due to the migration of free cations and anions in polymer electrolytes and, (ii) preferred site hopping conduction through the polymer hetero sites [67]. The high-frequency region is visible in pure polymer and polymer salt system while for another sample it falls outside the measured frequency range. The high-frequency region in both systems follows well known Jonscher's power law (JPL) which is general characteristics of an SSIC as given by eqn. 13

$$\begin{cases} \sigma_{ac} = \sigma_{dc}(1 + (\omega/\omega_h)^n) & (13\ a) \\ \qquad and \\ \sigma_{ac} = 2\sigma_{dc}\ when\ \omega = \omega_h & (13\ b) \end{cases}$$

$\sigma_{ac}$ and $\sigma_{dc}$ are the AC and DC conductivities of electrolyte, while A and n are the frequency independent Arrhenius constant and the power law exponent is dimensionless frequency exponent represents the degree of interaction between mobile ions and its surrounding, where $0 < n < 1$ ( the red solid line is for the fitting of JPL). The value of n is zero for an ideal Debye dielectric dipolar-type and 1 for an ideal ionic-type crystal. $\omega_h$ is hopping frequency and at this particular frequency ac conductivity becomes double of dc conductivity. For some polymers n may be greater than unity also [68-70].

But, one drawback associated with the JPL is that it is valid only at the high-frequency window and it does not consider the contribution of the universal electrode polarization (EP) region which appears in the low-frequency region. So, we investigate our results using the model proposed by the Roy et al., [71] for a better understanding of the ion dynamics parameters and of course the accuracy as it covers the whole frequency window. The effective complex conductivity can be written as

$$\sigma^*_{eff} = \left(\frac{1}{\sigma_b} + \frac{1}{i\omega C_{dl}}\right)^{-1} + i\omega C_b \quad (14)$$

Now, considering the equation 14 the real and imaginary part of the conductivity can be written as

$$\sigma'(\omega) = \frac{\sigma_b^2 C_{dl}\omega^\alpha \cos\left(\frac{\alpha\pi}{2}\right) + \sigma_b(C_{dl}\omega^\alpha)^2}{\sigma_b^2 + 2\sigma_b C_{dl}\omega^\alpha \cos\left(\frac{\alpha\pi}{2}\right) + (C_{dl}\omega^\alpha)^2} \quad (14\,a)$$

and,

$$\sigma''(\omega) = \frac{\sigma_b^2 C_{dl}\omega^\alpha \sin\left(\frac{\alpha\pi}{2}\right)}{\sigma_b^2 + 2\sigma_b C_{dl}\omega^\alpha \cos\left(\frac{\alpha\pi}{2}\right) + (C_{dl}\omega^\alpha)^2} + \omega C_b \quad (14\,b)$$

The real and imaginary part of conductivity in the high frequency region has been expressed as equation 15 a & b;

$$\begin{cases} \sigma'(\omega) = \sigma_b\left[1 + \left(\frac{\omega}{\omega_h}\right)^n\right] & (15\,a) \\ \quad\quad\quad and \\ \sigma''(\omega) = A\omega^s & (15\,b) \end{cases}$$

Here, all parameters have the same meaning as earlier and both 'n' & 's' have a value less than unity. Now, to investigate the complete frequency response we replace the $\sigma_b$ in Eq. 14 a by Eq. 15 a & Eq. 14 b by Eq. 15 b. This leads to the final equations which are used for further simulating the experimental data and are expressed as equation 16 a & b;

$$\sigma'(\omega) = \frac{\left(\sigma_b\left[1 + \left(\frac{\omega}{\omega_h}\right)^n\right]\right)^2 C_{dl}\omega^\alpha \cos\left(\frac{\alpha\pi}{2}\right) + \sigma_b\left[1 + \left(\frac{\omega}{\omega_h}\right)^n\right](C_{dl}\omega^\alpha)^2}{\left(\sigma_b\left[1 + \left(\frac{\omega}{\omega_h}\right)^n\right]\right)^2 + 2\sigma_b\left[1 + \left(\frac{\omega}{\omega_h}\right)^n\right]C_{dl}\omega^\alpha \cos\left(\frac{\alpha\pi}{2}\right) + (C_{dl}\omega^\alpha)^2} \quad (16\,a)$$

And

$$\sigma''(\omega) = \frac{(A\omega^s)^2 C_{dl}\omega^\alpha \sin\left(\frac{\alpha\pi}{2}\right)}{(A\omega^s)^2 + 2A\omega^s C_{dl}\omega^\alpha \cos\left(\frac{\alpha\pi}{2}\right) + (C_{dl}\omega^\alpha)^2} + \omega C_b \quad (16\,b)$$

Where, $C_{dl}$ is frequency independent double layer capacitance, $\omega$ is the angular frequency, s & $\alpha$ are exponent terms with value <1 and $C_b$ is the bulk capacitance of solid polymer electrolyte [66]. Figure 14 a shows the profile of the real part of conductivity and red solid lines are corresponding fits. At low frequency($\omega \to 0$) ion jump at faster rate from one coordinating site ($-\ddot{O}-$) to another for $\omega < \omega_h$ and this increases the relaxation time. But, at high frequency ($\omega \to \infty$) two competing hopping mechanism are known; one is unsuccessful hopping when ion jumps back to its initial position (correlated forward–backward–forward) and another is successful hopping when the neighborhood ions become relaxed with respect to the ion's position (the ions stay in the new site), i.e., successful hopping. For, $\omega > \omega_h$ the number of successful hopping is more and indicates a more dispersive ac conductivity [72-73]. When the frequency exceeds $\omega_h$, $\sigma_{ac}$ increases proportionally on, where n < 1. The simulated results provides an estimation of the dc conductivity, double layer capacitance, bulk capacitance and fractional exponent n and s as shown in Table 7. Value of n and s is less than unity which suggests that SPE system is a pure ionic conductor and the addition of the SN increases the dc conductivity and reduction in hopping time evidence the fast ion migration. The addition of SN may alter the polymer chain arrangement and increase in disorder or flexibility of the polymer chain results in faster segmental motion and is in correlation with the impedance, FTIR study and dielectric analysis as discussed in previous sections.

Now, the frequency dependence of imaginary part of the conductivity along with the contribution of electrode polarization (EP) can be written as in equation 14 b. The signature of frequency dependent imaginary part of ac conductivity ($\sigma$) for PEO-NaPF$_6$- *x* wt. SN (0, 1, 2, 3, 4, 5) is shown in Figure 14 b. The complete agreement between the experimental and fitted data in the measured frequency range strengthens the consistency of data. All plots show the low frequency electrode polarization region and intermediate frequency dc conductivity region followed by high-frequency dispersion region. The growth of polarization region is not in whole frequency window but starts at onset frequency ($\omega_{on}$) and a minimum in $\sigma''$ is noticed in all polymer electrolyte systems. This decrease of the $\sigma''$ is indication of the increase in the real part of permittivity ($\varepsilon'$) as in figure 8. Also above the $\omega_{on}$ the $\sigma's$ shows the frequency independent region as observed in the previous section. Further, with the decrease of frequency a peak in $\sigma''$ is growing up corresponding to a frequency $\omega_{max}$, where maximum electrode polarization is achieved. Again, the decrease in $\sigma''$ is evidenced [74-75]. The $\omega_{max}$ can also be correlated with the frequency corresponding to decreasing in the $\sigma'$ value. It is observed that with the addition of salt in the pure polymer matrix $\omega_{on}$ shifts toward the high frequency side window and indicates an increase in the frequency range of electrical polarization (EP). Further addition of SN in the polymer salt matrix shifts the $\omega_{on}$ toward the high frequency side and EP window becomes broad. Also, the onset of electrode polarization is associated with minima in the plot of the complex conductivity ($\sigma''$) and maxima in the loss tangent plot as depicted in Figure 15 (a & b). Then, after crossing the peak towards lower frequency, the imaginary part of conductivity $\sigma''$ again shows decreasing trend [76]. Thus, both the $\sigma'$ & $\sigma''$ variation are in good correlation with each other in terms of the onset frequency ($\omega_{on}$) and $\omega_{max}$.

### 3.8. Modulus study

The modulus study is extensively used for better insight into the dielectric behavior of the polymer electrolytes and can be utilized to investigate the conductivity relaxation by suppressing the electrode polarization effect at low frequency. So, the dielectric data is transformed into the modulus data and can be related to the permittivity by the following relation (equation 17)

$$M^* = M' + jM'' = \frac{1}{\varepsilon_*} ; \; M' = \frac{\varepsilon'}{\varepsilon'^2 + \varepsilon''^2} \; and \; M'' = \frac{\varepsilon''}{\varepsilon'^2 + \varepsilon''^2} \quad (17 \text{ a})$$

$$M^* = j\omega C_o Z^* = \omega C_o Z'' + j\omega C_o Z' \quad (17 \text{ b})$$

Figure 16 shows the plot of M´´ vs. M´ for different SN content and all show a deformed semicircle which indicates the presence of heterogeneous or broad relaxation processes. The presence of single semicircle is indication of single relaxation in polymer electrolyte system. Although complete semi-circle was not observed, extrapolation may be done for better visibility. The smallest semi-circle diameter for SPE 6 in Figure 16 e is associated with the highest capacitance. Figure 17 a & b depicts the real and imaginary parts of electric modulus versus frequency respectively and dispersion is seen in the high frequency region. From the graph, it is observed that $M' \rightarrow 0$ at low frequency and large capacitance is observed as indicated by the long tail is attributed to electrode polarization effect. In the high frequency window the increase in $M'$ is corresponds to dispersion in polymer electrolyte system and is observed when there is less restoring force for mobile charges on the application of field and this evidences the lone-range mobility of charge carriers [77]. The value of both real and imaginary part of modulus is independent of frequency indicating the null effect of the electric field in dipole orientation.

The lower value in modulus spectra (M″) as in Figure 17 b is an indication of transport of ion. From the figure, it is observed that M″ spectrum is approaching towards the relaxation at high-frequency side which is not within the experimental frequency range [36]. The modulus relaxation time ($\tau_m$) has been obtained and shows a decrease with the addition of SN in the polymer salt system. The low value of relaxation time indicates the fast cation migration from one coordinating site (ether group) of PEO to another site and it results in an increase in ionic conductivity of solid polymer electrolyte. Finally, after the detailed electrical and dielectric analysis, it becomes important to explore the transport parameters in detail to fulfill the criteria of polymer electrolyte for the desired application in energy storage/conversion devices. So, in the next section transport parameters are evaluated and are correlated with the previous analysis as discussed in the previous sections.

### 3.9. Transport Parameters Study

The fundamental transport parameters such as number density (n), ion mobility ($\mu$) and diffusion coefficient (D) are very crucial for completing the dielectric analysis of polymer electrolyte systems. The ionic conductivity is a desirable property for polymer electrolytes, it directly depends on the mobility and number of charge carriers. As in earlier results it was observed that the ionic conductivity increases with the addition of SN in the polymer salt system. This increase may be due to an increase of more number of free charge carriers due to high polarity of SN which supports the complete dissociation of salt, other due to mobility ($\mu$) and diffusion coefficient (D) of transport charge carriers (ions). Former one depicts the degree of ease of ions transport and later one indicates the passage of ions (due to the concentration gradient) through the medium on the application of external electrical field. Since, all the three parameters are dependent on each other and all together depends on the number of free ions numbers which is obtained by the deconvolution of FTIR pattern (as calculated above) [21, 78] and following equation 18 (a-c) are used to obtain all parameters;

$$\begin{cases} n = \dfrac{M \times N_A}{V_{Total}} \times \text{free ion area (\%)} & (18\ a) \\ \mu = \dfrac{\sigma}{ne} & (18\ b) \\ D = \dfrac{\mu k_B T}{e} & (18\ c) \end{cases}$$

In equation (18 a), M is the number of moles of salt used in each electrolyte, $N_A$ is Avogadro's number ($6.02 \times 10^{23}$ mol$^{-1}$), $V_{Total}$ is the total volume of the solid polymer electrolyte, and $\sigma$ is dc conductivity. In equation (18 b), 'e' is the electric charge ($1.602 \times 10^{-19}$ C), $k_B$ is the Boltzmann constant ($1.38 \times 10^{-23}$ J K$^{-1}$) and 'T' is the absolute temperature in equation 18 c.

Table 8 lists the values of $V_{Total}$, free ions (%), n, $\mu$, D obtained using the FTIR method [79]. From the table it is clearly observed that the number of free charge carriers increases with the addition of SN in the polymer salt system. This increase in a number of charge carriers is due to complete dissociation of salt in the polymer salt matrix. Further mobility also follows same behavior and highest mobility value for the SPE 6 system indicates highest ionic conductivity for the same system and also confirms our results obtained from the impedance study (Figure 18). As diffusion coefficient is inversely proportional to the viscosity so it may be said in another way that viscosity of polymer chain decrease with the addition of SN which enhances the migration of ion form one coordinating to another site. In, the next section it is correlated with the various relaxation times.

This increase in the mobility of an ion in the plasticized polymer electrolyte may be due to the enhancement of free volume or amorphous content which is an intrinsic requirement for faster ion dynamics. Also, the high dielectric constant of SN results in the release of more number of charge carriers up to an optimum concentration (4 wt. % SN) and followed by a decrease at a high SN content. This decrease in dielectric constant at high SN content may be due to the insulating or plastic effect of SN which lowers the number of charge carriers for conduction. This result is also in absolute agreement with the associated FTIR ion-ion interaction and conductivity studies, various relaxation times and dielectric analysis which strongly fulfills the criteria of polymer electrolyte material for application in energy storage/conversion devices. As we have confirmed the enhancement of ionic conductivity and its correlation with the dielectric analysis as well as transport parameters. So, it would be more interesting for us to further correlate the various relaxation times which were obtained from different plots for the same relaxation peak with the ionic conductivity as analyzed in the following section.

## 3.10. Correlation of ionic conductivity (σ) with double layer capacitance ($C_{dl}$), dielectric strength ($\Delta \varepsilon$) with various relaxation times ($\tau_\varepsilon, \tau_Z, \tau_{\tan \delta}, \tau_m, \tau_h$)

After exploring the crucial investigations in brief now, it becomes important to build a correlation between the various parameters that promote our idea of fast ionic transport in the investigated solid polymer electrolyte. These parameters are free ion carriers, dielectric strength, double layer capacitance, high ionic transference number and fundamental one is different relaxation times such as: $\tau_\varepsilon, \tau_Z, \tau_{\tan \delta}, \tau_m, \tau_h$. This report provides the detailed examination of dielectric parameters over various concentration of SN at room temperature and correlation is built between all of them for a better insight of the ion dynamics in the case of the polymer electrolytes. Figure 19 (a-h) shows the variation of different relaxation times ($\tau_{\varepsilon'}, \tau_{\tan \delta}, \tau_Z, \tau_m$), double layer capacitance and ionic conductivity with the varying concentration of SN. We, obtain here the one-to-one correspondence between ionic conductivity with the relaxation time obtained from different processes of the present investigated system. Figure 19 (a-c) depicts that the ionic conductive increases with the addition of SN in the polymer salt matrix and may be due to release of more free charge carriers which increase the double layer capacitance or polarization. Also, the increase in dielectric relaxation strength or the ionic polarization ($\Delta\varepsilon$) for the same SN content indicated complete dissociation of salt (Figure 19 c). While Figure (19 d-h) depicts the decrease in the relaxation time for the highly ionic conducting sample. Basically, relaxation time governs the segmental motion of polymer chain and the lower value is desirable for better performance of the electrolyte. It can be concluded that the decrease of relaxation time is accountable for the enhanced diffusion coefficient since diffusion coefficient is in inverse relation to the relaxation time. The decrease of relaxation times revels that the ion dynamics is strongly governed by the segmental motion of the polymer chain and increase the rate of segmental dynamics ($\tau_{\tan\delta}$) increases the ion mobility ($\tau_m$) in the faster relaxing media ($\tau_Z$) [80]. The hopping of ion is crucial finding here to support the ion migration in the investigated polymer electrolyte system and is supported by the decreased hopping time ($\tau_h$) as shown in figure 19 h. This reduction in the hopping time of the ion enables the faster jump of ion from one coordinating site to another and hence the high ionic conductivity. Also, the increase of relaxation time at a high content of SN evidence the decrease of ionic conductivity due to plasticization effect played by SN which increases the relaxation time.

Finally, we conclude that the above analysis of variation of various relaxation times is in absolute agreement with the ionic conductivity value and follows the same trend. This exploration of different relaxation times and their correlation is sufficient to briefly explore the ion dynamics using the dielectric analysis. Also, one remarkable point to be noted here is that all relaxation times of same relaxation process follows the order $\tau_{\varepsilon'} > \tau_{\tan\delta} > \tau_z > \tau_m$ as reported by W. Cao [54]. It means that the dielectric constant and tangent delta loss peak lies at a lower frequency while the impedance and modulus peak located at high frequency. So, due to their different relaxation peak position a relaxation peak may be seen in one process and absent in another as in modulus spectra shown in Figure 17. This agreement suggest the validity of the proposed new loss tangent equation and Sigma representation which agreed well with the traditional methods. So the detailed investigation in the various relaxation times and their correlation with the ionic conductivity value validate the genuineness of investigated polymer electrolyte system.

### 3.11. Correlation of hopping frequency with the polymer segmental motion

The dc conductivity of a polymer electrolyte system depends on the number density of charge carriers, ion mobility, and the ion charge; $\sigma_{dc} = \sum_i q_i n_i \mu_i$. Along with them, diffusion coefficient (D) plays an active role in the ion transport that varies with the hopping frequency. The ion transport in the investigated system is achieved via the coordinating sites provided by the host polymer having the electron-rich ether group ($-\ddot{O}-$). When the salt is added to the host polymer, it interacts with the polymer chain and disrupts the polymer chain arrangement that improves the polymer flexibility. Also, the number of free charges for migration increases due to better dissociation of the salt and is further improved by the addition of highly polar SN. This results in the hopping of ions via the coordinating sites and successful hopping of ion favor the high ionic conductivity. The hopping contribution is achieved only at a particular frequency called hopping frequency. The SN in the investigated system also has nitrile group which plays a crucial role towards the increase of interaction between the polymer-ion, polymer-ion-SN, and polymer-SN. The Overall effect in the polymer matrix is to make the hopping of ions faster that attained by the faster segmental motion of the polymer chain and is in direct correlation with the ionic conductivity and the diffusion coefficient.

The plot of figure 20 explores the strong correlation of the dc conductivity (Figure 20 a) and the diffusion coefficient (Figure 20 b) with the hopping frequency. As the hopping frequency increases with the addition of the salt and SN in the polymer matrix same effect is evidenced in the conductivity and the diffusion coefficient. Since average hopping length is inversely proportional to the hopping frequency ($<\lambda^2> \propto (1/\omega_h)$), therefore a decrease of average hopping length indicates the rapid ion migration [63]. This may be due to the disruption of the crystalline phase and enhanced amorphous content which is a principal requirement for the ion dynamics in the study of the polymer electrolytes. Thus, it can be concluded that the polymer segmental motion plays and active role in enhancing the ion transport and supports approach of fast solid state ionic conductor (SSIC).

### 3.12. Self-Proposed Scheme

During the rigorous analysis of the dielectric study and ion dynamics of SN doped polymer electrolyte system, the encouraging results built insight to propose a logical and convincing model in order to justify the findings. Figure 21 comprising of three stages processes describing the ion movement on salt doping in host polymer and role of SN in such polymer electrolyte system to complete the mechanism.

*Stage 1*: In stage 1, various polymer chains having ether group (shown in red color) are shown close to each other due to crystalline nature of the polymer. Now, the salt is dispersed in the polymer matrix them the salt gets dissociated in the cation ($Na^+$) and anion ($PF_6^-$) owing to the presence of coordinating interaction between the electron rich ether group and the cation (blue circle) while anion get attached to the methyl group ($CH_2$) of the polymer chain. This separates the ion pair into cation and anion as the individual role is played now by the both in the polymer matrix. As sodium has Lewis acid character while ether group has Lewis base character and this leads to the strong tendency of the cation coordination due to the Lewis acid-base interaction. Thus host polymer chain provides the path for ion migration via the coordinating sites provided by polymer host. The polymer salt complex is obtained and interchain separation is increased as noticeable in stage 1, also evidenced by XRD results.

*Stage 2*: It shows the general interaction between the cation coordinated polymer chain and nitrile group associated SN. The addition of SN in the polymer salt matrix modifies the chain arrangement. One important alteration which is noticed from the experimental results is an increase in the number of free charge carriers which is a desirable requirement for the fast SSIC. The SN having the nitrile group (brown color circle) in the structure helps in the salt dissociation. Now, the cation has two available sites for coordination one is ether group (-C=O) and the nitrile group (-C≡N). Also, the SN may get penetrated in between the polymer chains that diminishes the covalent bonding between the chains and chain sliding becomes smoother that leads to the faster ion transport. The increase of the free volume due to SN penetration indicates the enhancement of the amorphous content that is desirable for the fast ion dynamics system. This increase in free volume directly linked with the increased interchain separation that increases with the addition of SN in polymer salt complex as in XRD analysis. When there is a variation of SN content in the polymer salt matrix SN plays a different role in the matrix, It is explained in stage 3 of the below section.

But Now, *Stage 3*: (a) When the SN content is low then the covalent bonding is diminished as but partial salt dissociation observed there as evidence same by the FTIR deconvolution. Some ion pair is still in the polymer matrix (dotted red circle) and this makes less number of available free charge carriers for transport.

Stage 3: (b) when the SN content is increased, then at an optimum content salt get completely dissociated and the SN plays an active role in enhancing the ion transport due to chain modification. Here, nitrile group of the SN and the ether group of the polymer chain compete for the interaction with the cation. As oxygen have more electronegativity than the nitrogen so the cation has more chance to get coordinated with the ether group. The Guttmann number (GN) also plays a beneficial role. GN is more for the oxygen than the N so the cation has more possibility to make coordination bond with the ether group. So, the nitrile group helps in the release of the cation form the coordinating sites and makes the easier flow of the ion inside the polymer matrix by increasing the interchain separation (R) in the polymer matrix.

Stage 3: (c) when the SN content is higher than the optimized content then the negative role is played by the SN. As it has inherent plastic or insulating nature so this behavior dominates at a high content and the two possible reasons may be behind the reduction of ionic conductivity or increase of the relaxation time. One fundamental reason that may exist here is the blockage of a cation by the SN (indicated by a dotted red ellipse) or destruction of conductive paths. Another reason may be the decreased tendency of the chain sliding. In another word, we can say that the viscosity of polymer chain get increased and it becomes tougher for the cation to move due to the poor chain segmental motion. This was evidenced by the increase of various relaxation time owing to the slow segmental motion that directly indicates the reduction in the ionic conductivity or slower ion dynamics. Overall the proposed scheme is in ono-to-one correspondence with the electrical and dielectric parameters such as (dielectric constant and relaxation time) which validates the experimental findings.

## 4. Conclusions

In summary, the structural, electric and dielectric properties has been examined in polymer electrolyte system comprising of PEO-NaPF$_6$+SN prepared by solution cast technique. The XRD analysis confirms the decrease of crystallinity with the addition of SN. FTIR spectrum analysis confirms the presence of polymer-ion, ion-ion, interaction and complex phase formation. Impedance spectroscopy provides the enhancement of ionic conductivity with a maximum of 4 wt. % SN. The transference number has been evaluated ~0.99 invariably for all polymer electrolyte system which confirms the system is purely ionic in nature. The decrease in dielectric permittivity and dielectric loss with an increase in frequency is observed and high value of dielectric constant for the system having 4 wt. % SN is in good agreement with the impedance study. The loss tangent was plotted with frequency and maximum loss at a particular frequency confirms relaxation in polymer electrolytes and the peak of frequency in loss plot shifts toward the high-frequency side suggesting faster ion dynamics. The fitting dielectric parameters enables us to determine the dielectric strength, relaxation times etc. The sigma representation ($\sigma''\ vs.\ \sigma'$) approach explained clearly the suppressed region of Cole-Cole plot ($\varepsilon''\ vs.\ \varepsilon'$) at lower frequencies. The increase in the radius of semicircle indicates a decrease of relaxation time, hence favors high ionic conductivity also is in one-to-one correspondence with the above results. The frequency dependent real and imaginary conductivity were fitted in whole frequency window with corresponding equations. Various transport parameters such as: number density (n), ion mobility (μ) and diffusion coefficient (D) were evaluated to explore the ion dynamic study and are in good agreement with the other experimental data. Finally the various relaxation times were plotted together and all are in close agreement with the ionic conductivity data satisfying the order $\tau_{\varepsilon'} > \tau_{\tan\delta} > \tau_z > \tau_m$. The significant correlation observed between the ionic conductivity, transport parameters and relaxation times in the SN doped polymer salt complex system. A scheme is proposed which explores the various possible interactions between the constituents of the polymer composite matrix. It may thus anticipate that the reported study allows us to use the polymer electrolyte in the application in energy storage conversion device sector.

**Supporting Information**.

Supplementary discussion on molecular relaxation time and simulated results of Sigma representation. Supplementary figures S1 and S2 (PDF)

**Acknowledgement**

One of the authors (AA) thankfully acknowledges the Central University of Punjab, Bathinda for providing the fellowship and partial support from UGC start-up grant (GP-41). AA is thankful to the Gaurav Joshi and Sourav Kalra Research Scholar at the Central University of Punjab, Bathinda for support in the scheme presentation.

**Figure Caption**

**Figure 1**. The structures of PEO, NaPF$_6$ and SN.

**Figure 2**. XRD spectra of the (a) pure PEO, (b) PEO-NaPF$_6$ ; 0 wt. % SN), (c) 1 wt. % SN, (d) 2 wt. % SN, (e) 3 wt. % SN, (f) 4 wt. % SN and (g) 5 wt. % SN.

**Figure 3.** FESEM micrographs of the (a) PEO (b) PEO-NaPF$_6$ (c) PEO-NaPF$_6$-2 % SN (d) PEO-NaPF$_6$-3 % SN (e) PEO-NaPF$_6$-4 % SN and (f) PEO-NaPF$_6$-5 % SN polymer electrolytes.

**Figure 4**. FTIR spectra of pure PEO (SPE 1) and polymer salt (SPE 2) with SPE 3-7 with x wt. % of SN.

**Figure 5**. Deconvolution of the $PF_6^-$ vibration mode in the wavenumber range 800 cm$^{-1}$ to 900 cm$^{-1}$ for (a) 0 wt. % SN, (b) 1 wt. % SN, (c) 2 wt. % SN, (d) 3 wt. % SN, (e) 4 wt. % SN, (f) 5 wt. % SN.

**Figure 6.** Log-log plots of the complex impedance ($Z'' \: vs. \: Z'$) for the pure PEO and polymer salt complex with $x$ wt. % SN (x=1, 2, 3, 4, 5). The inset shows the graph of the highest conducting sample.

**Figure 7.** Variation of polarization current as a function of time for prepared solid polymer electrolyte at RT.

**Figure 8.** Cole-Cole plot for the a. PEO, polymer salt + x wt. % SN [b. 0 wt. % SN, c. 1 wt. % SN, d. 2 wt. % SN, e. 3 wt.% SN, f. 4 wt. % SN, g. 5 wt. % SN] and h. symbolic representation of Cole-Cole plot.

**Figure 9**. Frequency dependence of the dielectric constant ($\varepsilon'$) for the a. PEO, polymer salt + x wt. % SN [b. 0 wt. % SN, c. 1 wt. % SN, d. 2 wt. % SN, e. 3 wt.% SN, f. 4 wt. % SN, g. 5 wt. % SN] and h. polarization mechanism. Solid lines are the best fit to the experimental data (a-g).

**Figure 10**. Variation of the real part of permittivity at different frequencies.

**Figure 11**. Frequency dependence of the dielectric loss ($\varepsilon''$) for the a. PEO (inset), polymer salt + x wt. % SN [b. 0 wt. % SN, c. 1 wt. % SN, d. 2 wt. % SN, e. 3 wt.% SN, f. 4 wt. % SN, g. 5 wt. % SN]. Solid lines are the best fit to the experimental data (a-g).

**Figure 12**. Frequency dependence of the tangent delta loss (tan δ) for the a. PEO, polymer salt + x wt. % SN [b. 0 wt. % SN, c. 1 wt. % SN, d. 2 wt. % SN, e. 3 wt.% SN, f. 4 wt. % SN, g. 5 wt. % SN] and h. comparison of frequency shift with SN concentration. Solid lines are the best fit to the experimental data (a-g).

**Figure 13**. $\sigma'' \: vs. \: \sigma'$ plot for the a. PEO, polymer salt + x wt. % SN [b. 0 wt. % SN, c. 1 wt. % SN, d. 2 wt. % SN, e. 3 wt.% SN, f. 4 wt. % SN, g. 5 wt. % SN] and h. comparative plot together and i. change in radii of a semi-circle with SN.

**Figure 14. (a)** Frequency dependence of real part of complex conductivity ($\sigma'$). Solid lines are absolute fit to the equation 16 a, **(b)** Frequency dependence of Imaginary part of complex conductivity ($\sigma''$) at RT. Solid lines are absolute fit to the equation 16 b. The dotted line shows the change in onset frequency ($\omega_{on}$) with SN content.

**Figure 15.** The imaginary part of complex conductivity and the tangent delta plot for PEO-NaPF$_6$ with 4 wt. % SN.

**Figure 16.** Argand plot of M$''$ vs. M$'$ for different SN content in PS matrix at room temperature.

**Figure 17**. Frequency dependent real (M$'$) and imaginary part (M$''$) of modulus spectra for various SPEs at room temperature.

**Figure 18**. (a & b) Variation of free ion area and transport parameters; (b) number density 'n', (c) ion mobility $'\mu'$ and (d) diffusion coefficient 'D'.

**Figure 19**. Plot of ionic conductivity, double layer capacitance, dielectric strength and various relaxation times ($\tau_{\varepsilon'}, \tau_{\tan\delta}, \tau_Z, \tau_m, \tau_h$) for the solid polymer electrolyte SPE (1-7).

**Figure 20.** The plot of the (a) dc conductivity and (b) diffusion coefficient for various solid polymer electrolyte against the hopping frequency.

**Figure 21.** Proposed scheme for the cation transport, *Stage 1*: Formation of coordination bond between the ether group of PEO and cation (Na$^+$), *Stage 2*: Interaction between the SN, host polymer and the salt, *Stage 3*: a. Low SN content, b. Optimized SN content, c. High SN content.

**Table Caption**

**Table 1**. Values of 2θ (degree), d-spacing (Å) and R (Å) of PEO-NaPF$_6$ with *x* wt. % SN (1, 2, 3, 4, 5) for (120) diffraction peak.

**Table 2.** The peak position of deconvoluted free ion and ion pair peak of SPE films.

**Table 3**. Different contributions of electrical conductivity and transference number for PEO-NaPF$_6$+ *x* wt. % SN.

**Table 4.** The fitted **ε'** ($\varepsilon_\infty, \Delta\varepsilon, \tau_{\varepsilon'}, \alpha$) and **ε''** ($\Delta\varepsilon, \tau_{\varepsilon''}, \alpha$) parameters at room temperature.

**Table 5.** The evaluated parameters (r, τ, α. $\tau_{\tan\delta}, \tau_m$) from the loss tangent plot at room temperature.

**Table 6.** The determined parameter, $\sigma_o$, $\sigma_\infty$, $\delta$ and $r$ from the plot of $\sigma''$ vs. $\sigma'$ for various SPEs at RT

**Table 7.** Comparison of fitted parameters evaluated from the simulated real and imaginary part of the complex conductivity for different SPEs at RT.

**Table 8**. The values of V$_{\text{Total}}$, free ions (%), n, $\mu$, D obtained using the FTIR method.

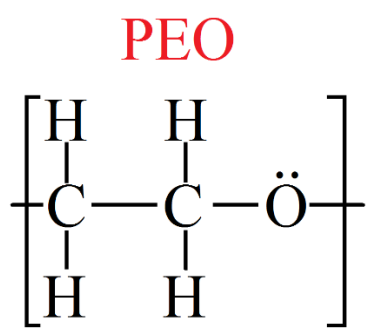 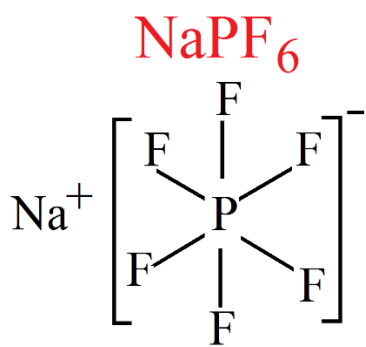 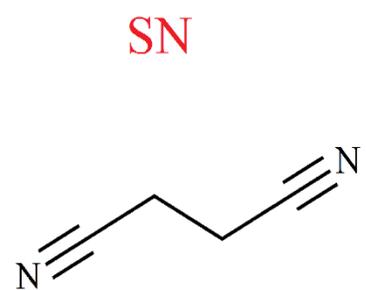

**Figure 1**

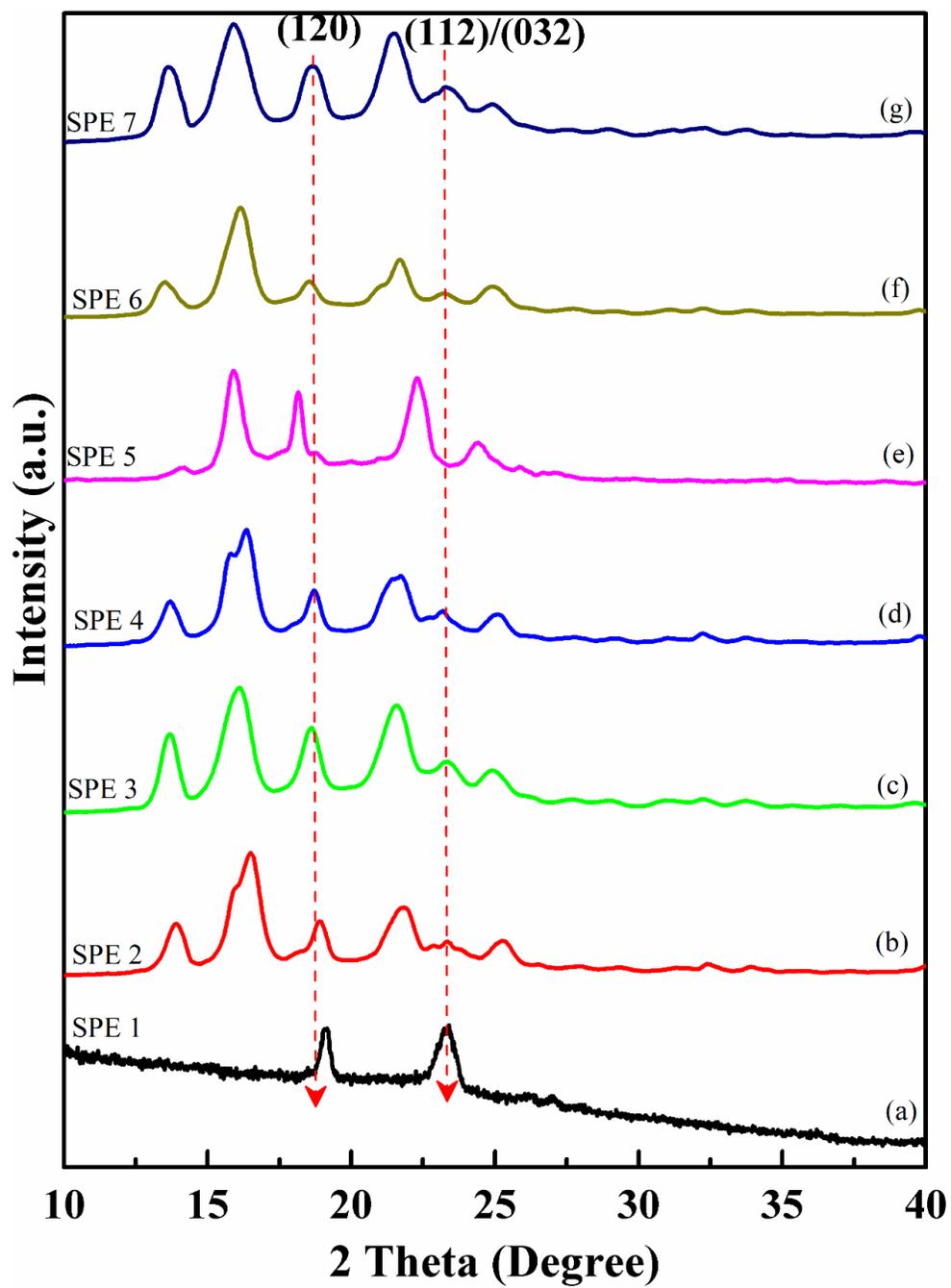

**Figure 2**

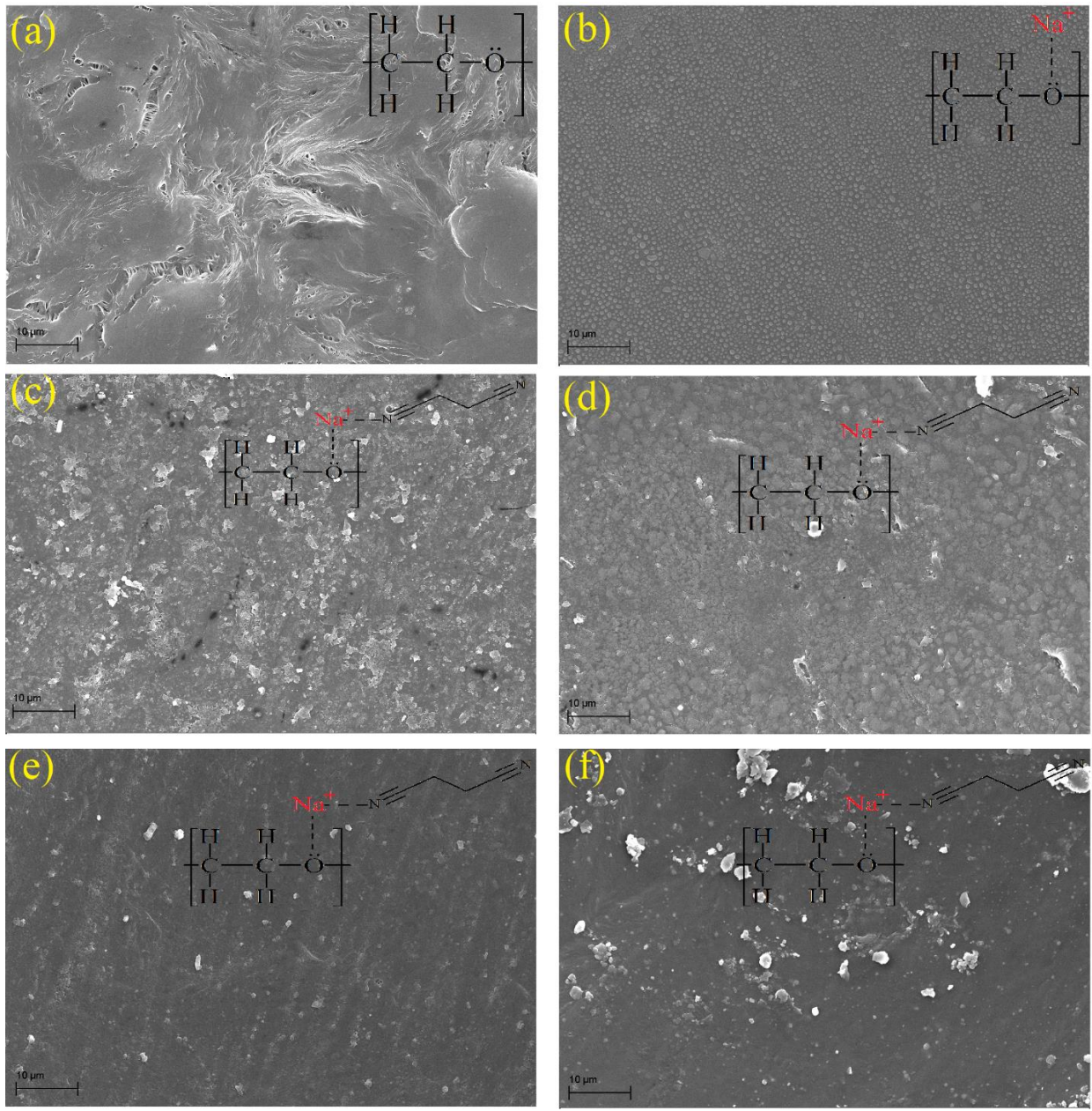

**Figure 3**

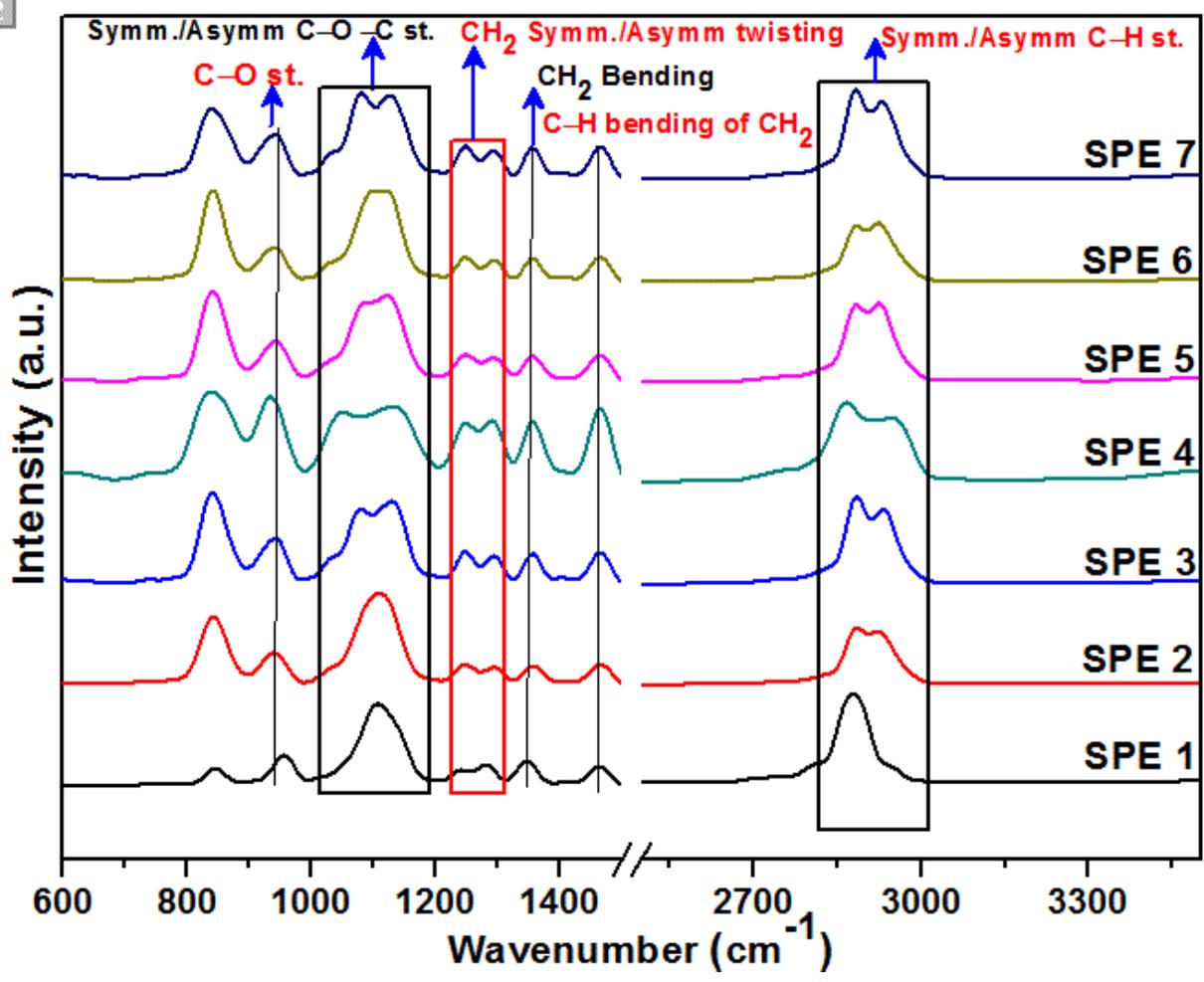

**Figure 4**

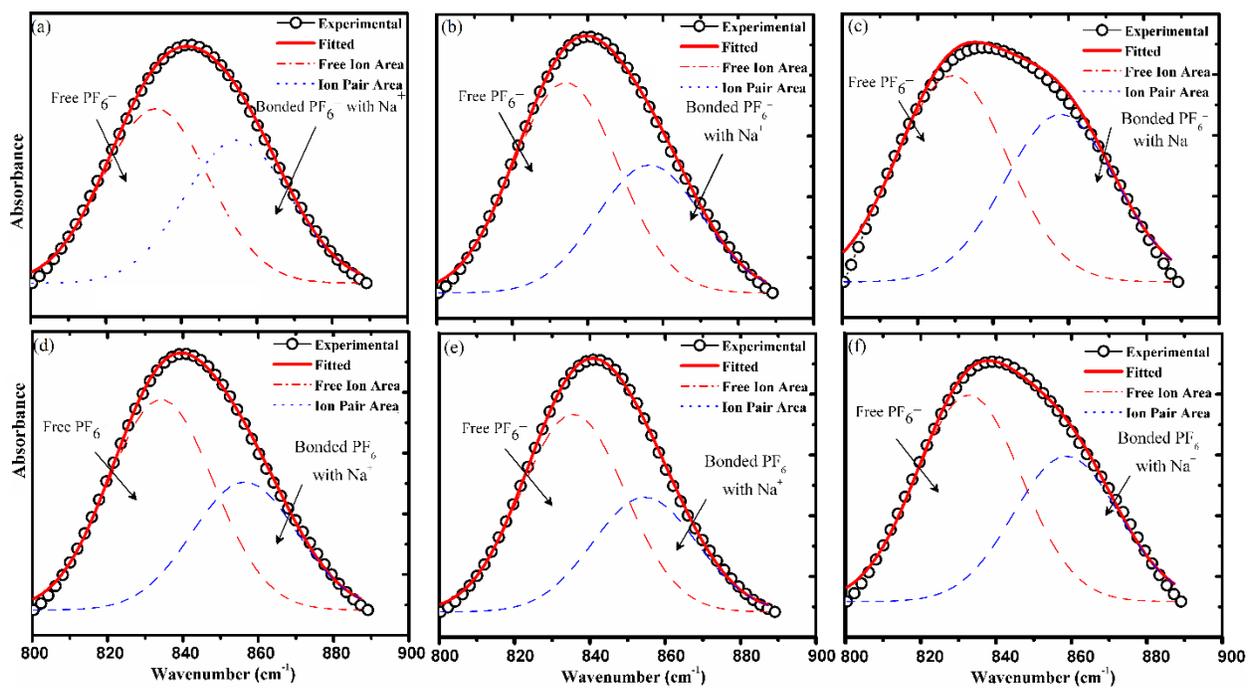

**Figure 5**

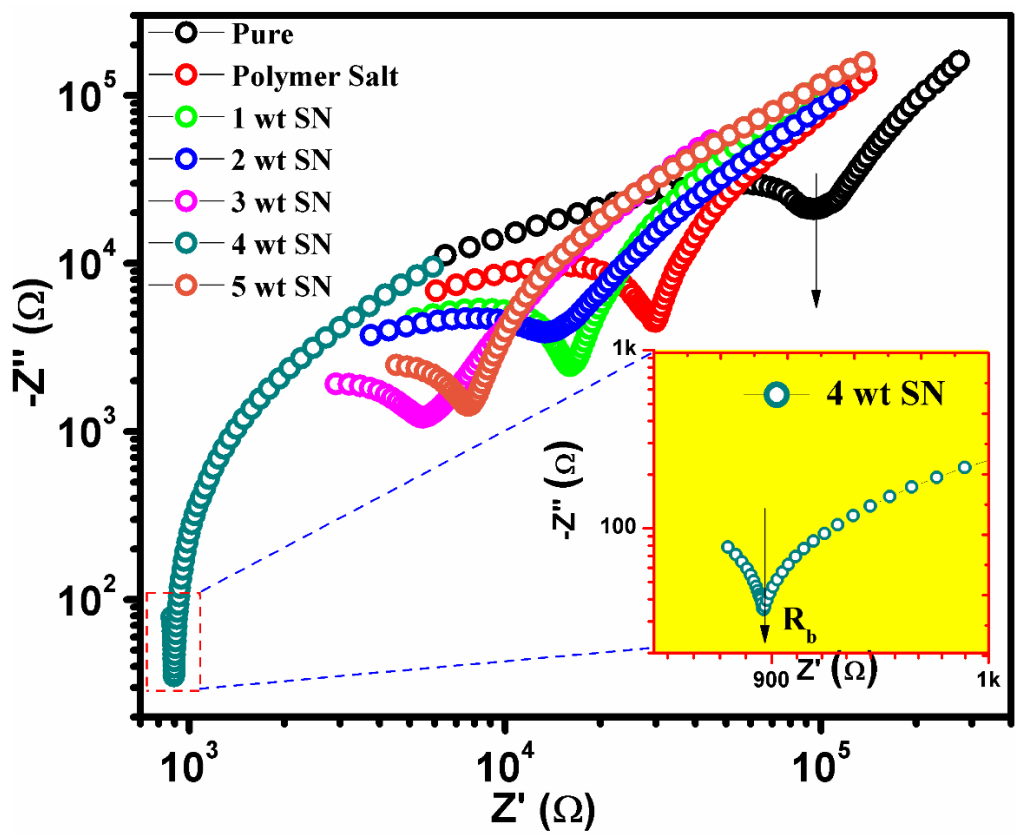

**Figure 6**

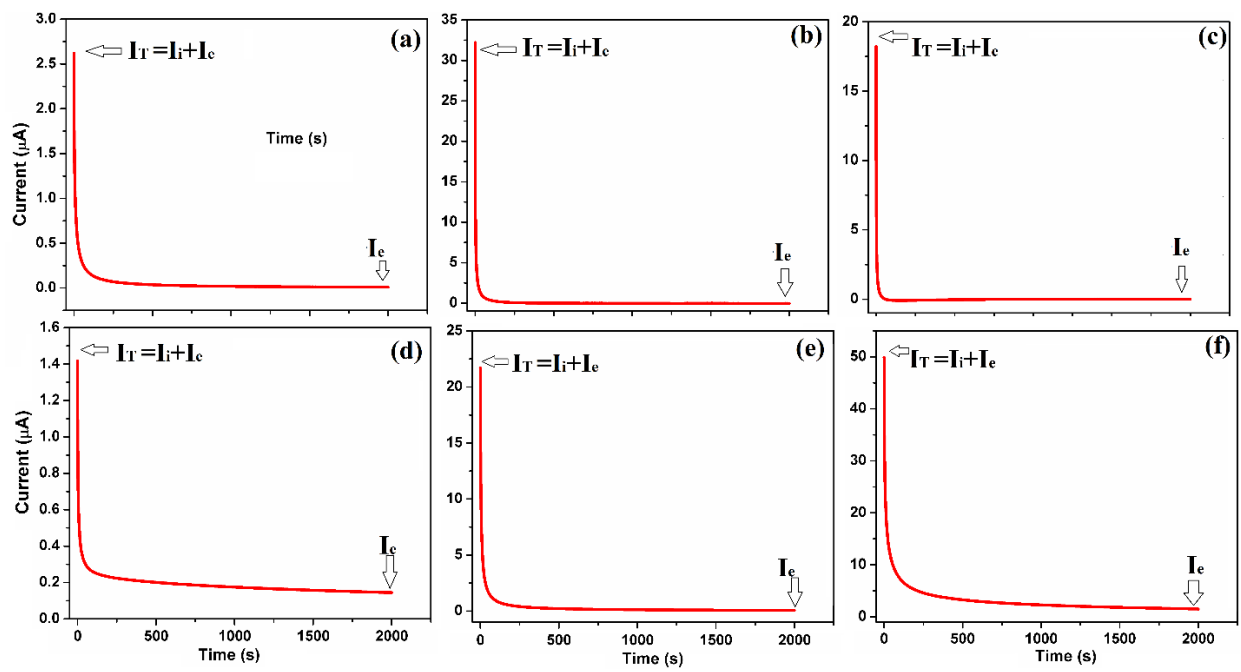

**Figure 7.**

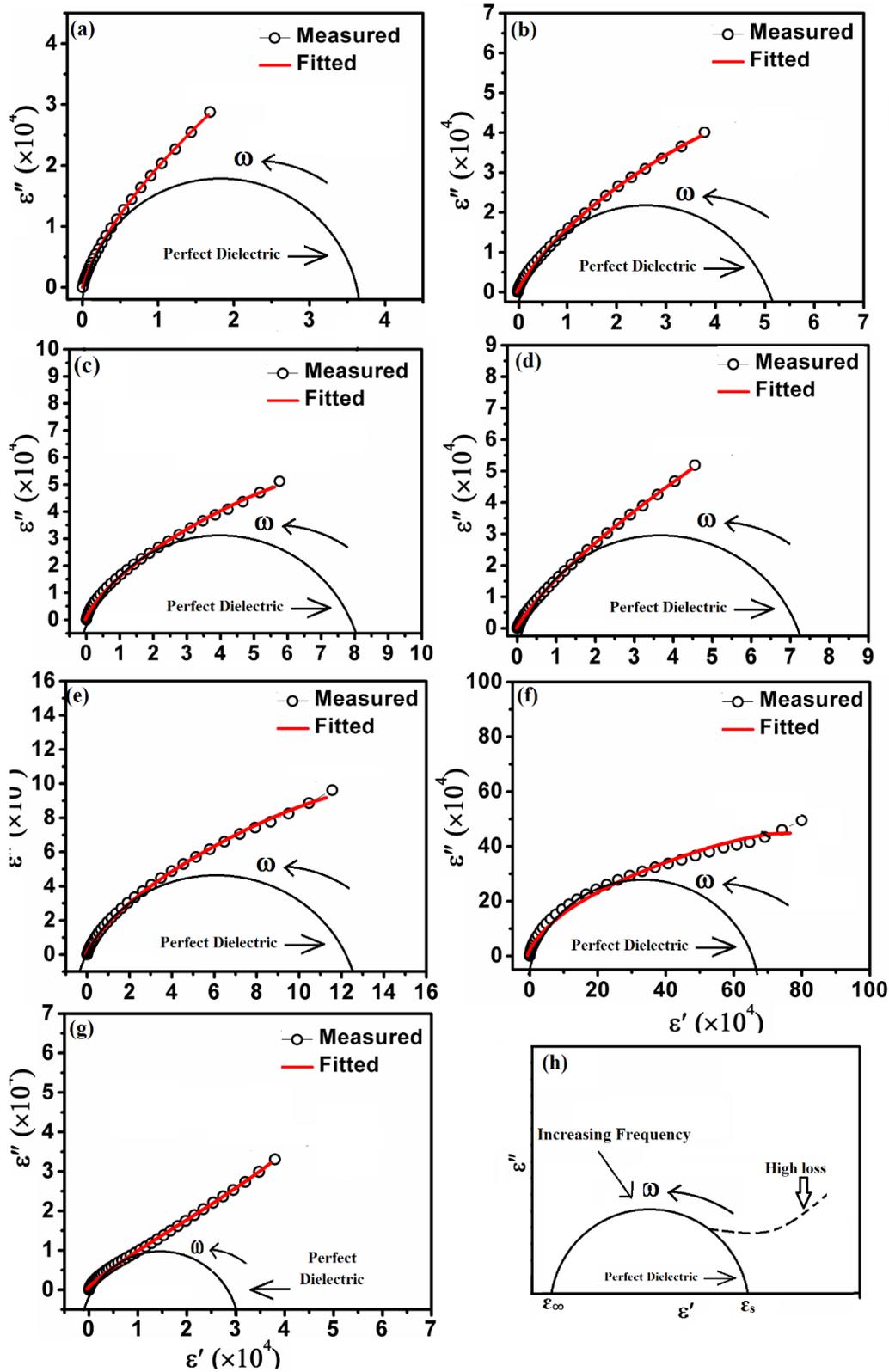

**Figure 8.**

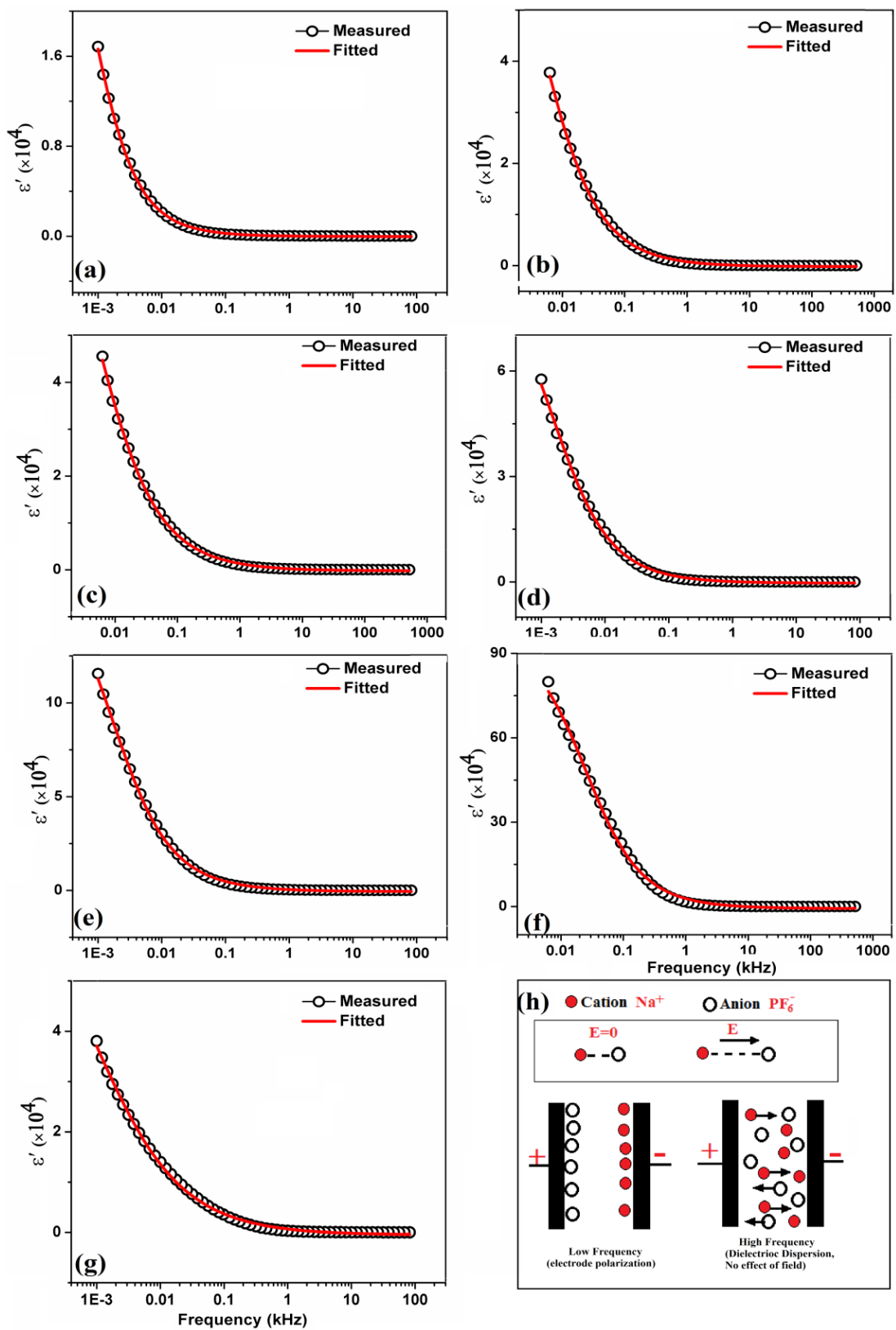

**Figure 9.**

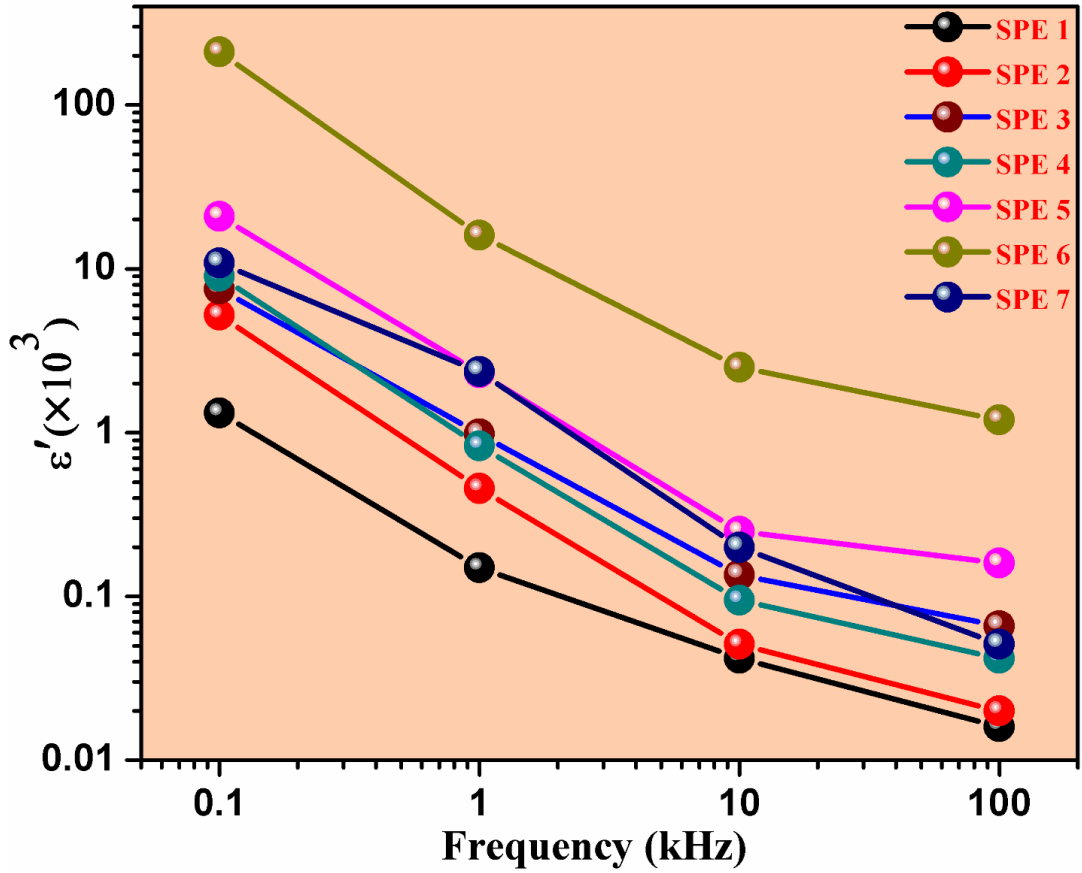

Figure 10.

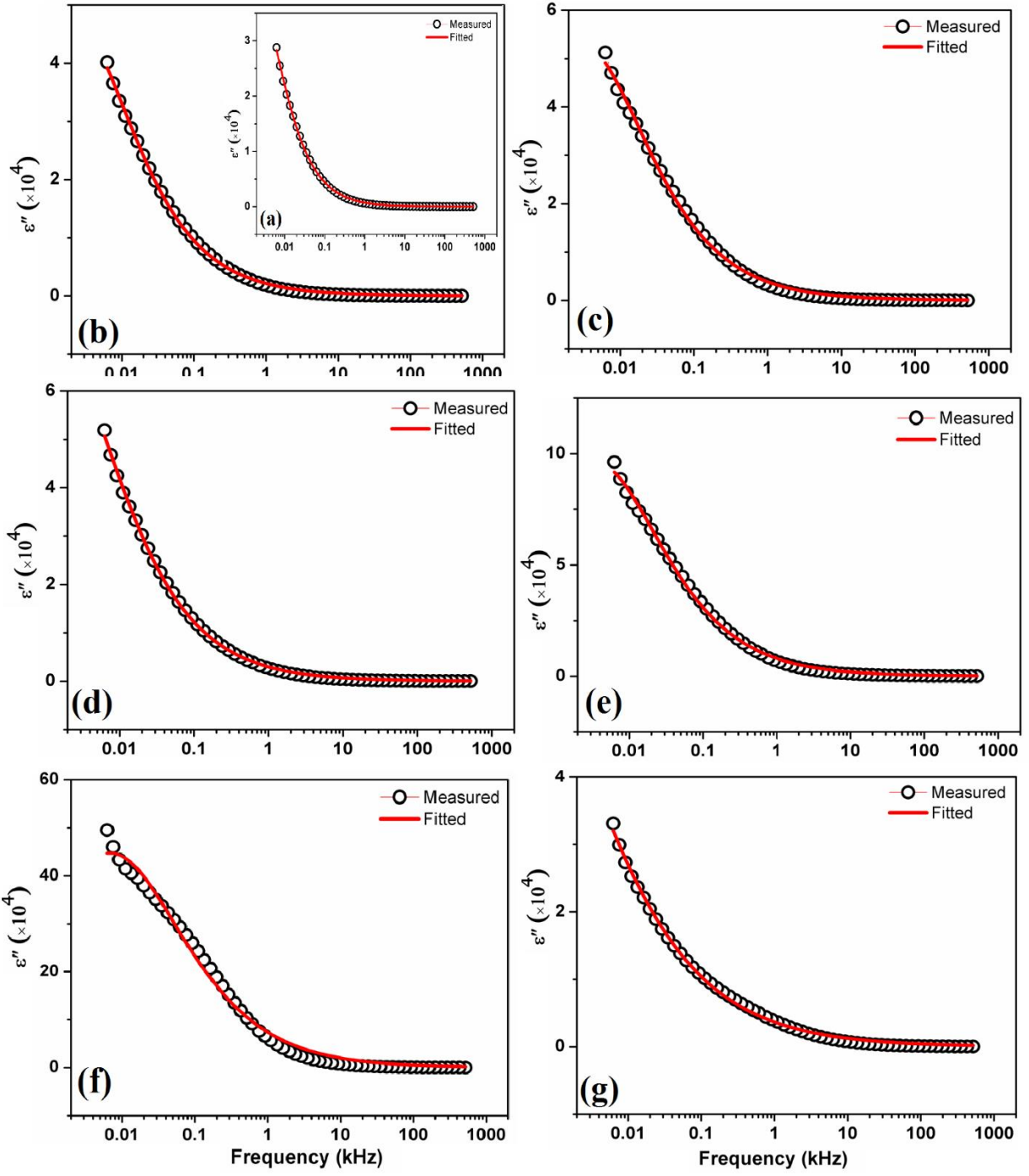

**Figure 11.**

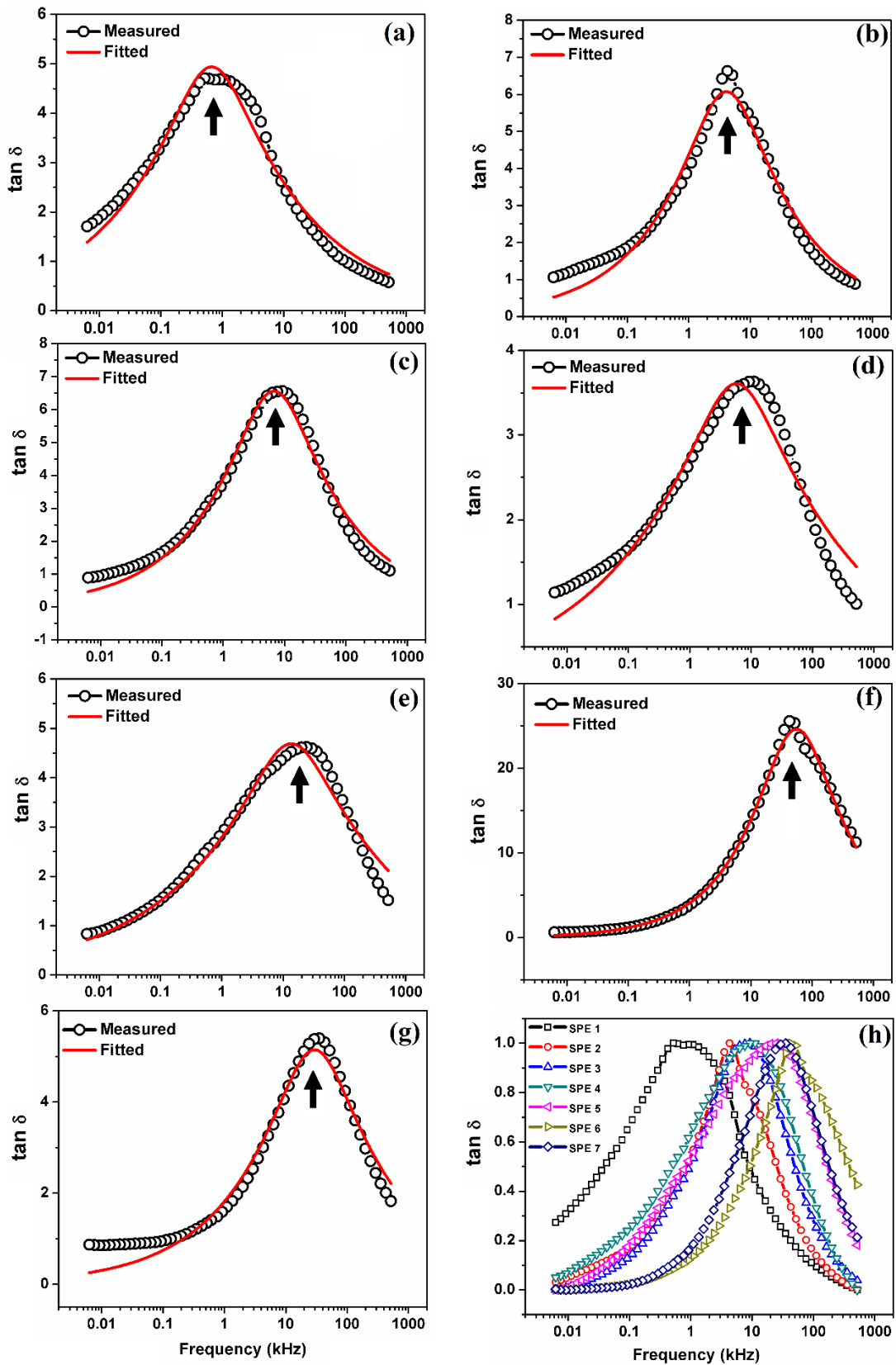

**Figure 12.**

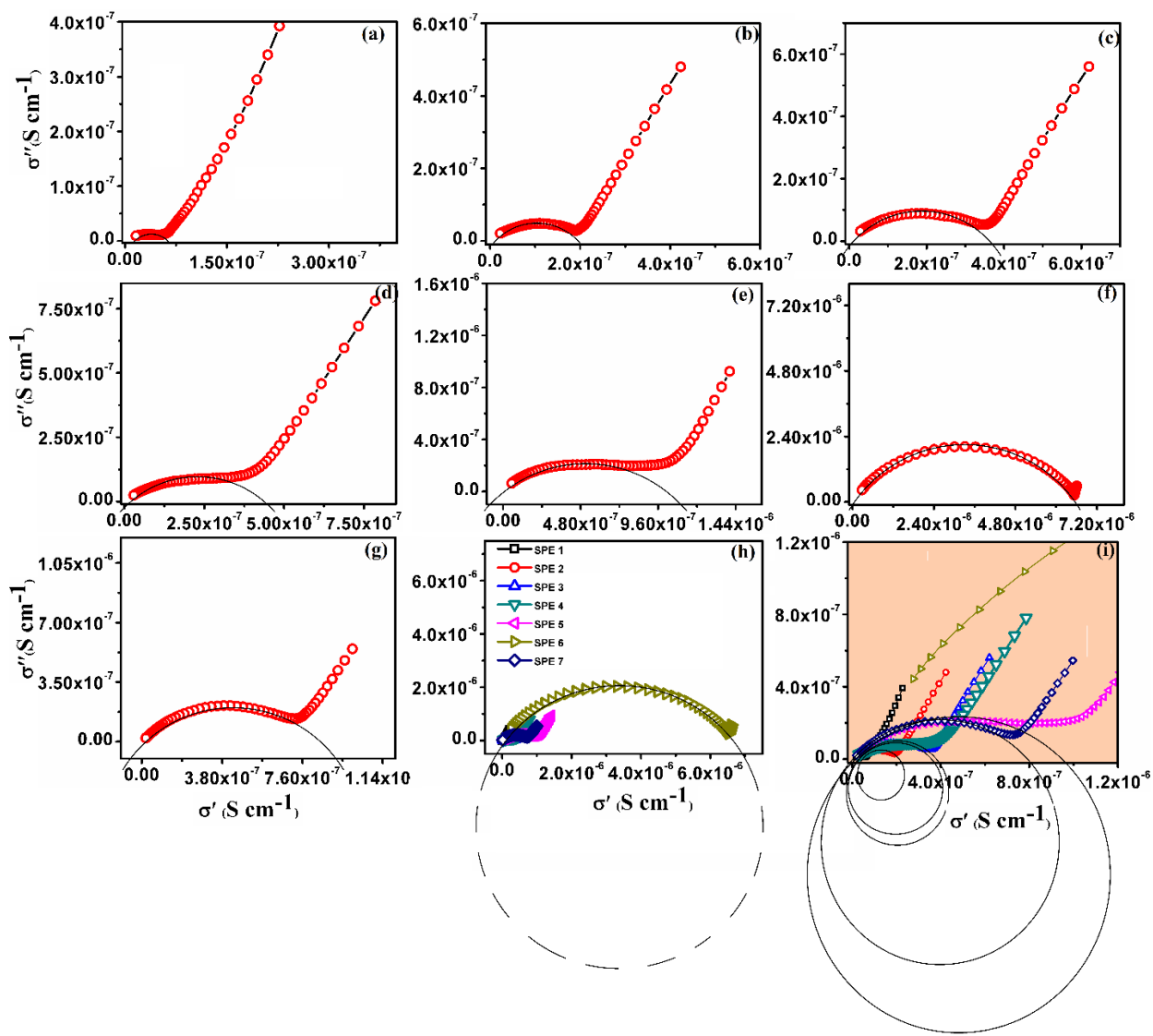

**Figure 13.**

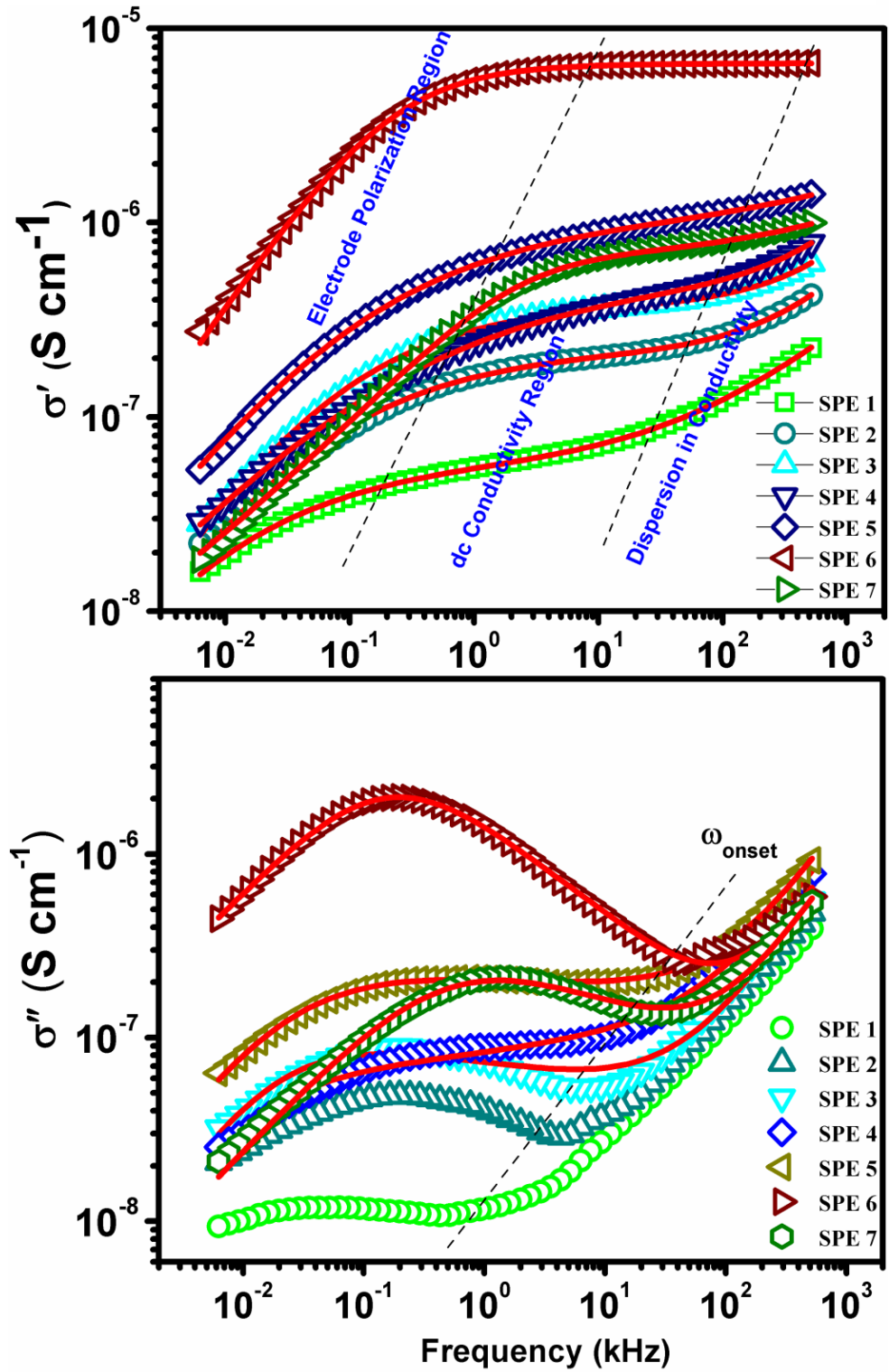

**Figure 14.**

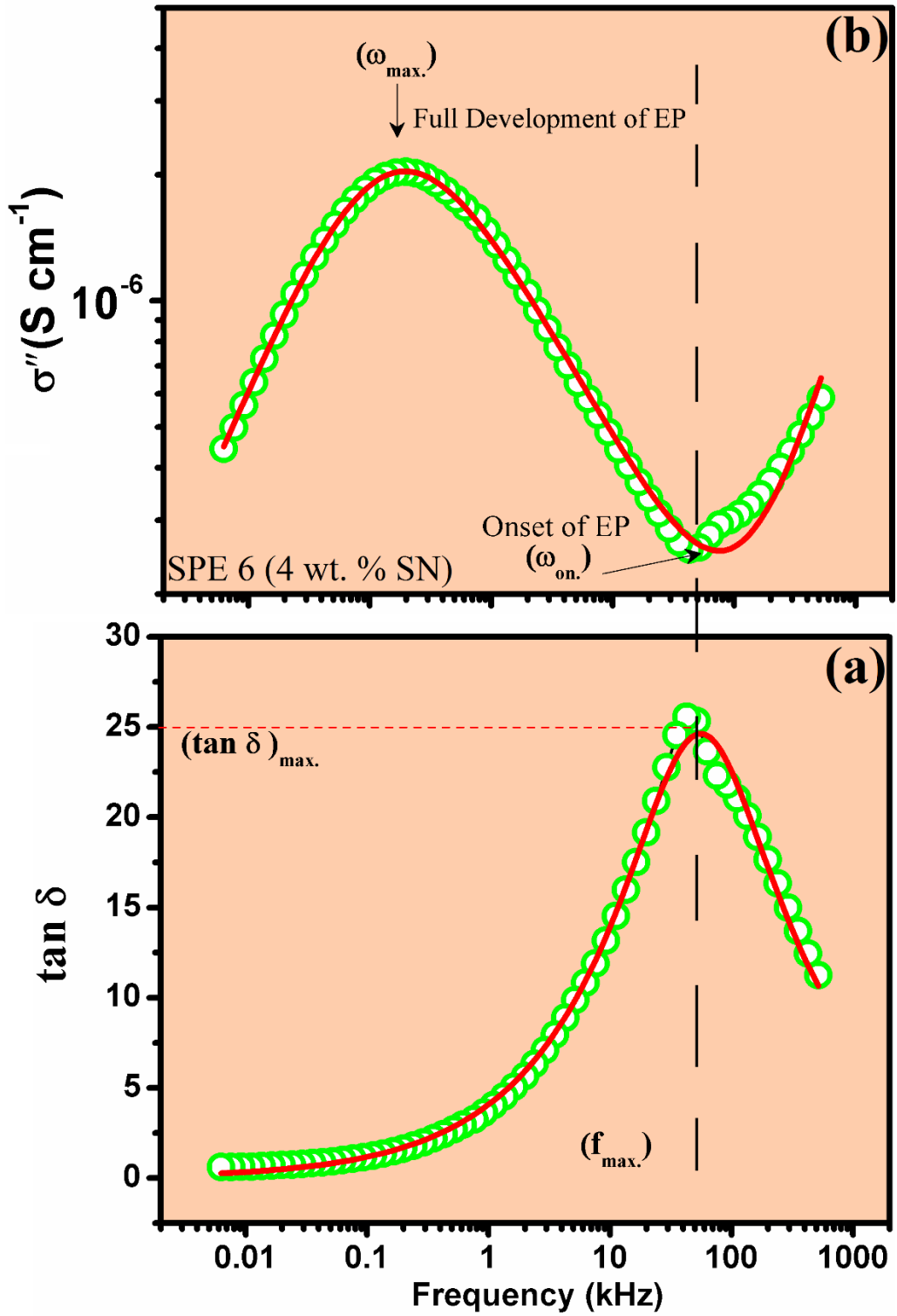

**Figure 15.**

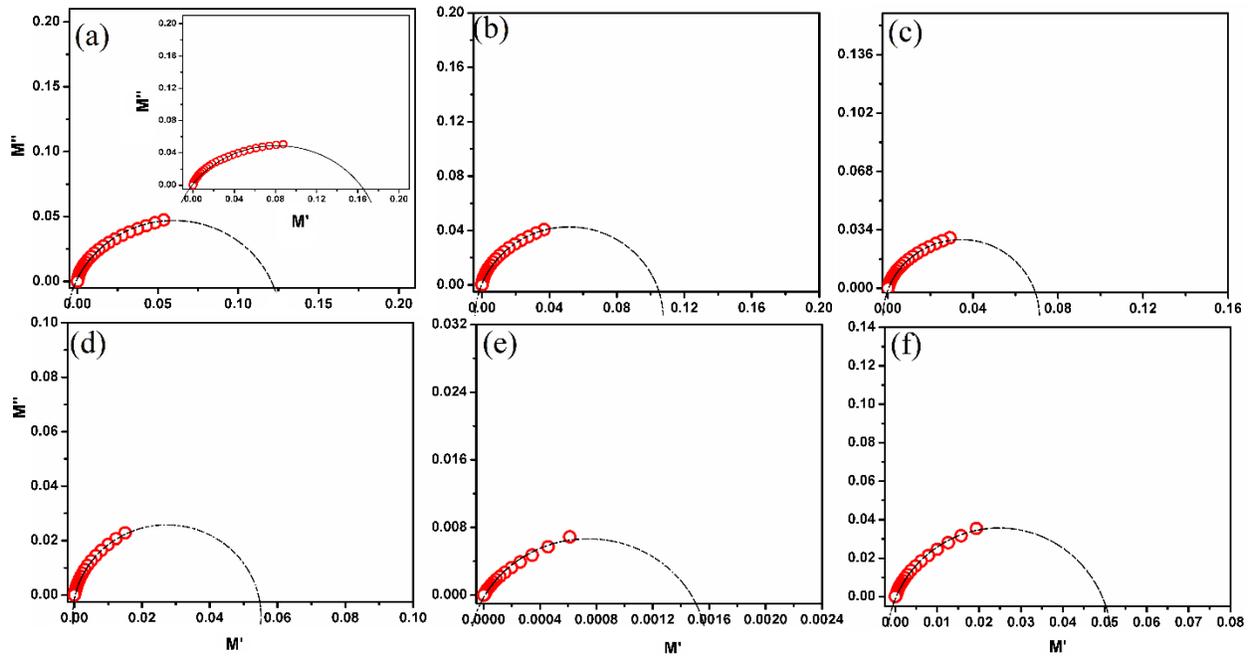

**Figure 16.**

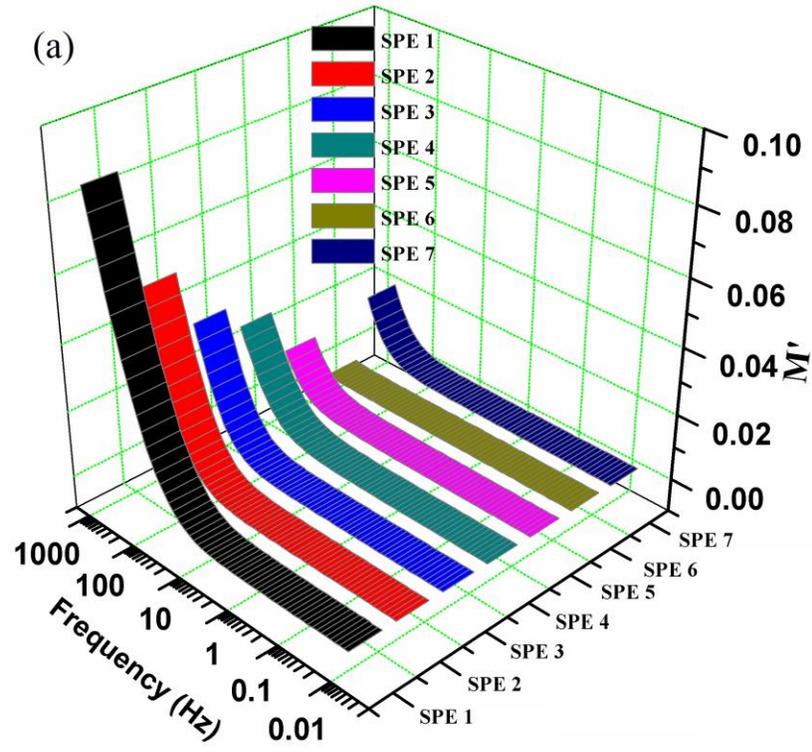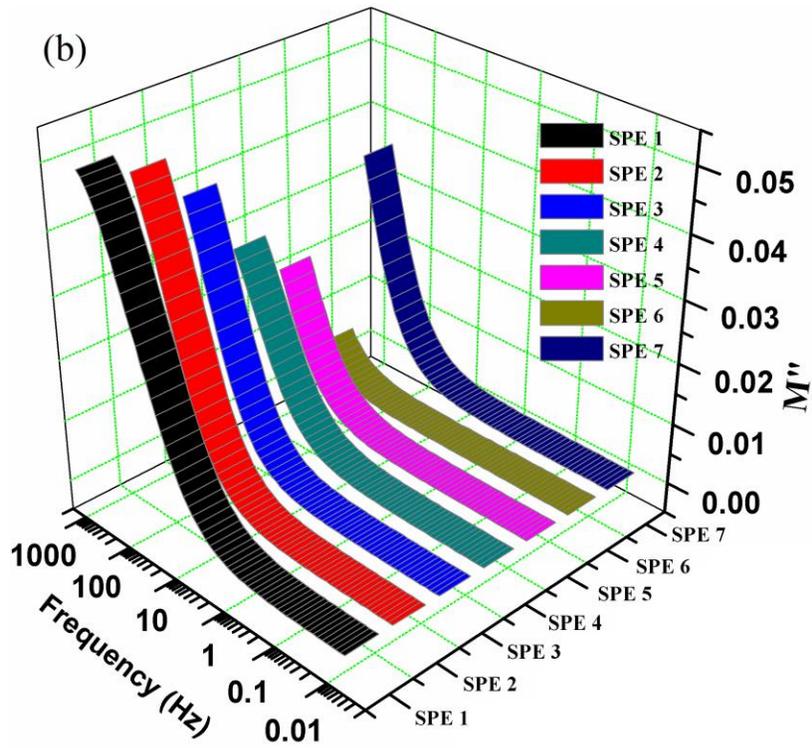

**Figure 17.**

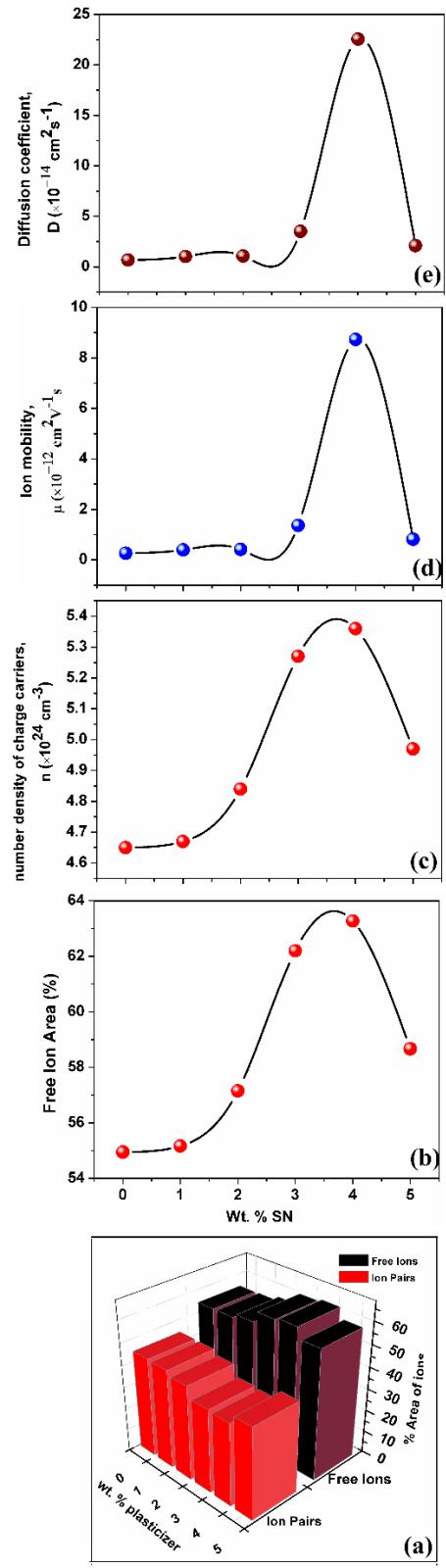

**Figure 18.**

**Figure 19.**

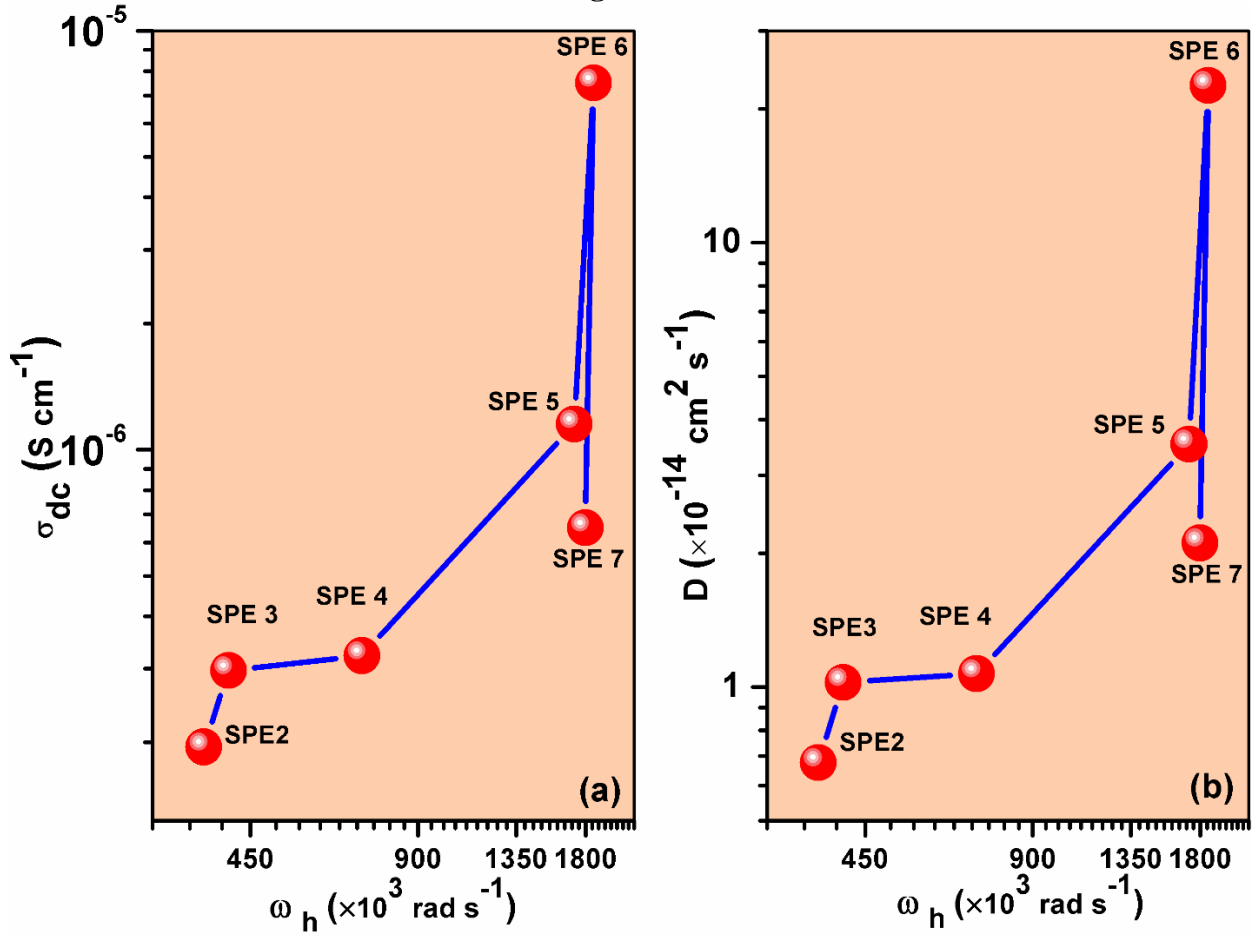

**Figure 20.**

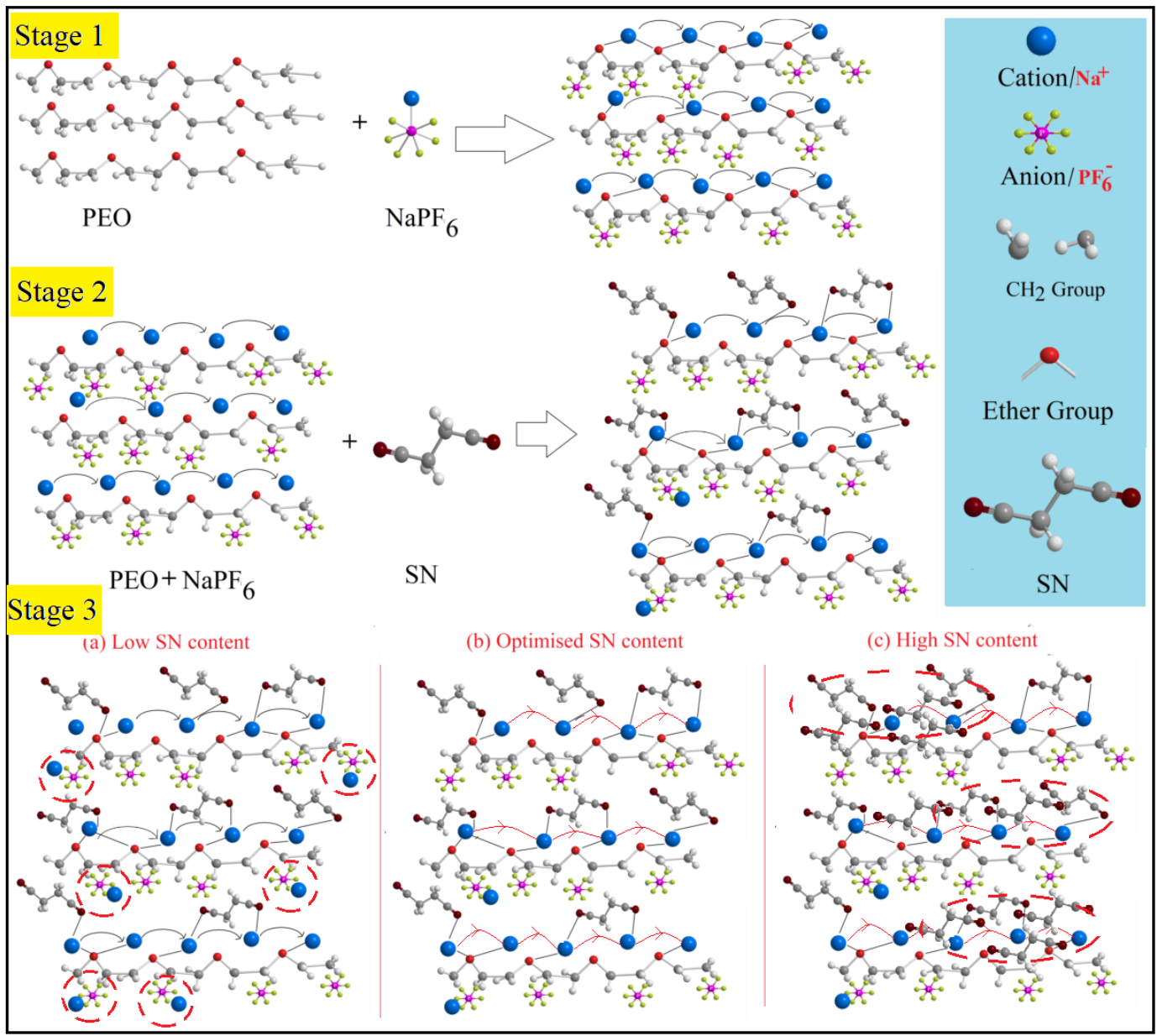

**Figure 21.**

Table 1. Values of 2θ (degree), d-spacing (Å) and R (Å) of PEO-NaPF$_6$ with *x* wt. % SN (1, 2, 3, 4, 5) for (120) diffraction peak.

| Sample | 2θ (degree) | d-spacing (Å) | R (Å) |
|---|---|---|---|
| SPE 1 | 19.03 | 4.66 | 5.82 |
| SPE 2 | 18.87 | 4.69 | 5.87 |
| SPE 3 | 18.63 | 4.75 | 5.94 |
| SPE 4 | 18.38 | 4.82 | 6.02 |
| SPE 5 | 18.16 | 4.87 | 6.09 |
| SPE 6 | 18.54 | 4.78 | 5.97 |
| SPE 7 | 18.66 | 4.74 | 5.93 |

Table 2. The peak position of deconvoluted free ion and ion pair peak of SPE films.

| Sample Code | Free Ion | | Ion Pair | | Corr. Coff. ($r^2$) |
|---|---|---|---|---|---|
| | Area (%) | Wavenumber (cm$^{-1}$) | Area (%) | Wavenumber (cm$^{-1}$) | |
| *SPE 2* | 54.95 | 833 | 45.04 | 854 | 0.99 |
| *SPE 3* | 55.17 | 830 | 44.82 | 857 | 0.98 |
| *SPE 4* | 57.15 | 833 | 42.84 | 854 | 0.99 |
| *SPE 5* | 62.20 | 834 | 37.79 | 856 | 0.99 |
| *SPE 6* | 63.27 | 835 | 37.32 | 854 | 0.99 |
| *SPE 7* | 58.67 | 833 | 41.32 | 858 | 0.99 |

Table 3. Different contributions of electrical conductivity and transport number for PEO-NaPF$_6$+

| Sample Code | Transference Number | Electrical Conductivity (S cm$^{-1}$) | Electronic Conductivity (S cm$^{-1}$) | Ionic Conductivity (S cm$^{-1}$) |
|---|---|---|---|---|
| SPE 1 | - | 4.95×10$^{-8}$ | - | - |
| SPE 2 | 0.90 | 1.95×10$^{-7}$ | 1.95×10$^{-8}$ | 1.75×10$^{-7}$ |
| SPE 3 | 0.99 | 2.97×10$^{-7}$ | 2.97×10$^{-9}$ | 2.94×10$^{-7}$ |
| SPE 4 | 0.99 | 3.22×10$^{-7}$ | 3.22×10$^{-9}$ | 3.18×10$^{-7}$ |
| SPE 5 | 0.95 | 1.15×10$^{-6}$ | 9.75×10$^{-8}$ | 1.09×10$^{-6}$ |
| SPE 6 | 0.96 | 0.75×10$^{-5}$ | 4.52×10$^{-7}$ | 0.72×10$^{-5}$ |
| SPE 7 | 0.96 | 0.65×10$^{-6}$ | 3.9×10$^{-8}$ | 0.62×10$^{-6}$ |

*x* wt. % SN.

Table 4. The fitted $\varepsilon'$ ($\varepsilon_\infty$, $\Delta\varepsilon$, $\tau_{\varepsilon'}$, $\alpha$) and $\varepsilon''$ ($\Delta\varepsilon$, $\tau_{\varepsilon''}$, $\alpha$) parameters at room temperature

| Sample Code | $\varepsilon'$ | | | | $\varepsilon''$ | | |
|---|---|---|---|---|---|---|---|
| | $\varepsilon_\infty$ | $\Delta\varepsilon$ (×10³) | $\tau_{\varepsilon'}$ (s) | $\alpha$ | $\Delta\varepsilon$ (×10³) | $\tau_{\varepsilon''}$ (s) | $\alpha$ |
| *SPE 1* | -28.46 | 40.04 | 1.32 | 0.78 | 110.91 | 0.55 | 0.78 |
| *SPE 2* | -199.18 | 83.91 | 1.20 | 0.69 | 161.81 | 0.49 | 0.67 |
| *SPE 3* | -223.29 | 100.48 | 1.20 | 0.62 | 193.41 | 0.34 | 0.63 |
| *SPE 4* | -346.09 | 93.12 | 0.63 | 0.70 | 261.00 | 0.84 | 0.63 |
| *SPE 5* | -604.04 | 181.69 | 0.58 | 0.69 | 356.82 | 0.29 | 0.63 |
| *SPE 6* | -7094.61 | 976.84 | 0.03 | 0.70 | 1814.38 | 0.14 | 0.58 |
| *SPE 7* | -532.23 | 75.99 | 1.03 | 0.53 | 349.45 | 0.41 | 0.47 |

Table 5. The evaluated parameters (r, $\tau$, $\alpha$, $\tau_{\tan\delta}$, $\tau_m$) from the loss tangent plot at room temperature.

| Sample Code | r (×10⁴) | $\tau$ | $\alpha$ | $\tau_{\tan\delta}$ (×10⁻⁴ s) | $\tau_m$ (×10⁻⁶ s) |
|---|---|---|---|---|---|
| SPE 1 | 9.01 | 0.44 | 0.31 | 14.83 | 4.94 |
| SPE 2 | 2.04 | 0.03 | 0.42 | 2.41 | 1.68 |
| SPE 3 | 3.06 | 0.02 | 0.42 | 1.48 | 0.84 |
| SPE 4 | 17.45 | 0.07 | 0.24 | 1.73 | 0.41 |
| SPE 5 | 32.85 | 0.04 | 0.26 | 0.73 | 0.11 |
| SPE 6 | 56.38 | 0.01 | 0.54 | 0.17 | 0.02 |
| SPE 7 | 1.89 | 0.00 | 0.38 | 0.33 | 0.24 |

Table 6. The determined parameter, $\sigma_o$, $\sigma_\infty$, $\delta$ and $r$ from the plot of $\sigma''$ $vs.$ $\sigma'$ for various SPEs at RT

| Sample Code | $\sigma_o$ (S cm⁻¹) | $\sigma_\infty$ (S cm⁻¹) | $\delta$ (×10⁻⁷) | $r$ (×10⁻⁷) |
|---|---|---|---|---|
| SPE 1 | 6.12×10⁻⁹ | 7.03×10⁻⁸ | 0.64 | 0.32 |
| SPE 2 | 1.59×10⁻⁸ | 2.01×10⁻⁷ | 1.85 | 0.92 |
| SPE 3 | 2.27×10⁻⁸ | 3.59×10⁻⁷ | 3.36 | 1.68 |
| SPE 4 | 3.13×10⁻⁸ | 4.03×10⁻⁷ | 3.71 | 1.85 |
| SPE 5 | 6.31×10⁻⁸ | 1.11×10⁻⁶ | 11.0 | 5.51 |
| SPE 6 | 2.55×10⁻⁷ | 6.57×10⁻⁶ | 63.1 | 31.5 |
| SPE 7 | 1.81×10⁻⁸ | 7.98×10⁻⁷ | 7.79 | 3.89 |

Table 7. Comparison of fitted parameters evaluated from the simulated real and imaginary part of the complex conductivity for different SPEs at RT.

| The real part of complex conductivity (σ') [Eq. 16 a] | | | | | | |
|---|---|---|---|---|---|---|
| Sample Code | $\sigma_b$ (×10$^{-6}$ S cm$^{-1}$) | $\omega_h$ (×10$^5$) | $\alpha$ | n | $C_{dl}$ (µF) | $\tau_h$ (µs) |
| SPE 1 | 0.05 | 0.52 | 0.57 | 0.48 | 0.02 | 19.1 |
| SPE 2 | 0.22 | 3.71 | 0.85 | 0.47 | 0.03 | 2.69 |
| SPE 3 | 0.39 | 4.10 | 0.93 | 0.50 | 0.04 | 2.43 |
| SPE 4 | 0.52 | 7.13 | 0.87 | 0.36 | 0.07 | 1.40 |
| SPE 5 | 1.02 | 17.18 | 0.45 | 0.43 | 0.13 | 0.58 |
| SPE 6 | 91.4 | 186.04 | 0.84 | 0.41 | 33.1 | 0.53 |
| SPE 7 | 0.97 | 17.99 | 0.86 | 0.43 | 0.58 | 0.55 |
| The Imaginary part of complex conductivity (σ'') [Eq. 16 b] | | | | | | |
| Sample Code | A (×10$^{-6}$ S cm$^{-1}$) | s | $C_{dl}$ (µF) | $\alpha$ | $C_{dl}$ (pF) | |
| SPE 3 | 0.01 | 0.10 | 0.006 | 0.97 | 1.08 | |
| SPE 4 | 0.05 | 0.12 | 0.395 | 0.77 | 1.30 | |
| SPE 5 | 0.10 | 0.04 | 0.014 | 0.88 | 1.52 | |
| SPE 6 | 59.8 | 0.51 | 0.151 | 0.67 | 1.14 | |
| SPE 7 | 2.28 | 0.28 | 0.005 | 0.69 | 1.00 | |

Table 8. The values of $V_{Total}$, free ions (%), n, $\mu$, D obtained using the FTIR method.

| Sample Code | $V_{Total}$ ($\times 10^{-1}\ cm^3$) | Free Ions Area (%) | n ($\times 10^{24}\ cm^{-3}$) | $\mu$ ($\times 10^{-12}\ cm^2 V^{-1} s$) | D ($\times 10^{-14}\ cm^2 s^{-1}$) |
|---|---|---|---|---|---|
| SPE 2 | 0.2 | 54.95 | 4.65 | 0.26 | 0.67 |
| SPE 3 | 0.2 | 55.17 | 4.67 | 0.39 | 1.02 |
| SPE 4 | 0.2 | 57.15 | 4.84 | 0.41 | 1.07 |
| SPE 5 | 0.2 | 62.20 | 5.27 | 1.36 | 3.51 |
| SPE 6 | 0.2 | 63.27 | 5.36 | 8.73 | 22.5 |
| SPE 7 | 0.2 | 58.67 | 4.97 | 0.81 | 2.10 |